\documentstyle[11pt,aaspp4,flushrt]{article}
\received{}
\revised{}
\accepted{}
\ccc{}
\cpright{}{}

\newcommand{\kms}{km~s$^{-1}$}

\newcommand{\etal}{{\it et al.\/}}
\newcommand{\teff}{$T_{eff}$}
\newcommand{\grav}{log($g$)}

\newcommand{\ciso}{C$^{12}$/C$^{13}$}

\begin{document}

\title{C and N Abundances in Stars At the Base of the Red Giant 
Branch in M5
\altaffilmark{1} }

\author{Judith G. Cohen\altaffilmark{2},
Michael M. Briley\altaffilmark{3} and Peter B. Stetson\altaffilmark{4,5,6}}

\altaffiltext{1}{Based on observations obtained at the
W.M. Keck Observatory, which is operated jointly by the California
Institute of Technology, the University of California and the National
Aeronautics and Space Administration}

\altaffiltext{2}{Palomar Observatory, Mail Stop 105-24,
California Institute of Technology, Pasadena, California 91125
(jlc@astro.caltech.edu)}

\altaffiltext{3}{Department of Physics and Astronomy, 
University of Wisconsin Oshkosh,
Oshkosh, Wisconsin (mike@maxwell.phys.uwosh.edu)}

\altaffiltext{4}{Dominion Astrophysical Observatory, 5071  West 
Saanich Road,
Victoria, British Columbia   V9E 2E7 Canada (Peter.Stetson@hia.nrc.ca)}

\altaffiltext{5}{Guest User,  Canadian Astronomy Data Centre, 
which is operated by the
Herzberg Institute of Astrophysics, National Research Council of Canada}

\altaffiltext{6}{Guest Investigator at the UK Astronomy Data Centre}

\begin{abstract}
We present an analysis of a large sample of
moderate resolution Keck LRIS spectra of subgiant
($V \sim 17.2$)
and fainter stars in the Galactic globular cluster M5 (NGC 5904) 
with the
goal of deriving C and N abundances. Star-to-star
stochastic variations with significant range
in both [C/Fe] and [N/Fe] are found at all luminosities
extending to the bottom of the RGB at $M_V \sim+3$.
Similar variations in CH appear to be 
present in the
main sequence turnoff spectra, but the signal in the 
current sample is too low for a  detailed analysis.
The variations seen among the M5 subgiants are
consistent with the abundances found earlier by
Briley \etal\ (1992) for brighter giants in this cluster.
There is thus no sign of a change in the behavior of C and N
with evolutionary stage over the full range in luminosity
of the RGB and SGB, although a systematic decrease with
luminosity in the mean [C/H] smaller than
a factor of 2 cannot be ruled out with confidence at present.

The C and N abundances appear strongly anti-correlated, as
would be expected from the CN-cycle processing of stellar material.
Yet the present stars are considerably fainter than
the RGB bump, the point at which deep
mixing is believed to set in.  On this basis,
while the observed abundance pattern is consistent with 
proton capture nucleosynthesis,  we infer that
the site of the reactions is likely not
within the present sample, but rather in a population of more massive
(2 -- 5$M_{\odot}$) now defunct stars.

The range of variation
of the N abundances is very large and the sum of C+N increases
as C decreases. 
To reproduce this requires  
the incorporation not only of CN but also of ON-processed material.
Furthermore, the existence of this correlation is quite
difficult to reproduce with an external mechanism 
such as ``pollution'' with  material
processed in a more massive AGB star, which 
mechanism is fundamentally
stochastic in nature.  We therefore suggest
that although the internal mixing hypothesis has serious flaws,
new theoretical insights are needed and it should not be ruled 
out yet.

\end{abstract}


\section{Introduction\label{intro}}

By virtue of their large populations of coeval stars, the Galactic
globular clusters present us with a unique laboratory for the study of
the evolution of low mass stars.  The combination of their extreme
ages, compositions and dynamics also allows us a glimpse at the early
history of the Milky Way and the processes operating during its
formation. These aspects become even more significant in the context of
the star-to-star light element inhomogeneities found among red giants
in every cluster studied to date. The large differences in the surface
abundances of C, N, O, and often Na, Mg, and Al have defied a
comprehensive explanation in the three decades since their discovery.

Proposed origins of the inhomogeneities typically break down into two
scenarios: 1) As C, N, O, Na, Mg, and Al are related to proton capture
processes at CN and CNO-burning temperatures, material cycled through a
region in the upper layers of
the H-burning shell in evolving cluster giants may be
brought to the surface with accompanying changes in composition. While
standard models of low mass stars do not predict this ``deep mixing'',
several theoretical mechanisms have been proposed (e.g., the meridional
mixing of Sweigart \& Mengel 1979, and turbulent diffusion, 
Charbonnel 1994, 1995) with varying degrees of success. 
Moreover, there is ample
observational evidence that deep mixing {\it{does}} take place 
during the red
giant branch (RGB) ascent of metal-poor cluster stars (see the
reviews of Kraft 1994 and
Pinsonneault 1997 and references therein). 2) It has also become
apparent that at least some component of these abundance variations
must be in place before some cluster stars reach the giant branch.
Spectroscopic observations of main sequence turn-off stars in 47 Tuc
(Hesser 1978; Hesser \& Bell 1980; Bell, Hesser, \& Cannon 1983;
Briley, Hesser, \& Bell 1991; Briley \etal 1994, 1996; Cannon \etal\
1998) and NGC 6752 (Suntzeff \& Smith 1991, Gratton \etal\ 2000),
as well as our own work in M71 (Cohen 1999, Briley \& Cohen 2001,
Ram\'{\i}rez \& Cohen 2002)
have shown variations in CN
and CH-band and Na line strengths consistent with patterns found among
the evolved giants of these clusters. The assumption that these low
mass cluster stars are incapable of both deep dredge-up and significant
CNO nucleosynthesis while on the main sequence leads to the possibility
that the early cluster material was at least partially inhomogeneous in
these elements or that some form of modification of these elements has
taken place within the cluster. Suggested culprits include mass-loss
from intermediate mass asymptotic giant branch stars and supernovae
ejecta (see Cannon \etal\ 1998 for an excellent discussion of these
possibilities).

Thus the observed light element inhomogeneities imply that there is
some aspect of the structure of the evolving cluster giants which
remains poorly understood (the deep mixing mechanism), that the early
proto-clusters may have been far less homogeneous, that intermediate
mass stars may have played a greater role in setting the composition of
the present day low mass stars than previously thought, etc. Indeed,
the evidence cited in the reviews above has led many investigators to
suggest that a combination of processes is responsible, i.e., many
clusters contain star-to-star inhomogeneities established early in
their histories which have
subsequently been further altered by deep mixing during
the ascent of the RGB.
This of course greatly exacerbates the difficulty of achieving
an understanding of these issues, as a knowledge
of the composition of the more easily observed bright red giants will
not tell the whole story of their chemical history - one must also
understand the makeup of the main sequence stars.

In the present paper, we continue our earlier work on M71
by exploring the CH and CN
band strengths in a sample of low luminosity stars in the somewhat
more metal poor globular cluster M5. We adopt current values
from the on-line database of Harris (1996)
for the apparent distance modulus of M5 at $V$ of 14.31 mag
with a reddening of E(B--V) = 0.03 mag.  
Recent CCD photometric studies of this cluster,
focusing primarily on its age,
are given by  Johnson \& Bolte (1998) and Stetson \etal\ (1999).
Sandquist \etal\ (1996) discuss the predominantly blue
horizontal branch of M5.  We adopt the metallicity
[Fe/H] = $-$1.21 dex found by
Ivans \etal\ (2001) in a high dispersion abundance analysis of a large
sample of stars on the upper giant branch of M5.

We describe the sample
in \S\ref{section_phot} and \ref{section_spec}.  We outline our
measurement of the molecular band indices and their interpretation
in terms of the scatter in \S\ref{section_indices}.  With an 
assumption about
the O abundance, these are converted into C and N abundances,
from which we find a strong anti-correlation between C and N in
\S\ref{section_cnabund}.
A discussion of our results together with a
comparison with the trends seen among the red giants in M5
and in other globular clusters is
given in \S\ref{section_othergc} and \ref{section_discussion}.
A brief summary concludes the paper.

\section{Photometric Database \label{section_phot}}

The photometry of M5 employed here was carried out as part of a larger
program to provide homogeneous photometry for star clusters and nearby
resolved galaxies (Stetson, Hesser \& Smecker-Hane 1998,
Stetson 2000).  In the present instance, the photometry is
based on data obtained during 18 observing runs between 1984 and 
1998.  Of
these, some observing runs were carried out by PBS, data from other runs
were donated by collaborators (M.~Bolte, J.~Hesser, R.~McClure,
N.~Suntzeff), and data from still other runs were obtained through the
services of the Isaac Newton Group Archive.  Data origins include the 4m
and 0.9m telescopes on Kitt Peak and Cerro Tololo, the
Canada-France-Hawaii Telescope, the Isaac Newton Telescope, and the
Jacobus Kapteyn Telescope.
\footnote{Although some Keck images are available, very short
exposure times of only 1 or 2 seconds were used, and the
photometric accuracy of the data is then compromised at the
level of 2\% due to
issues involving the shutter open/close time and its uniformity
over the field.  In addition, the field of the available Keck images
is small.}  In all, some 374 images of parts of M5 in
filters corresponding to the {\it BVI\/} bandpasses were analyzed.  
Since
different field sizes and different field centers were available 
from the
various observing runs, no individual star actually appeared in 
more than
226 of those images.

Employing only data obtained under photometric conditions on nights when
numerous observations of fundamental and secondary standards (see 
Stetson
2000) were also obtained, we defined a network of 649 local photometric
standards in the M5 field.  A star was considered a ``local 
standard'' in
a given photometric bandpass only if it had been measured in that 
bandpass
under photometric conditions on at least five occasions, if the standard
error of the mean magnitude in that bandpass was less than 0.02$\,$mag,
{\it and\/} if the star showed no evidence for intrinsic variability
greater than 0.05$\,$mag, root-mean-square, in excess of random 
measuring
errors when all bandpasses were considered together.  Once this 
network of
local standards had been created, it was possible to use them to
calibrate the zero points of frames taken under non-photometric 
conditions
(the color terms of the photometric transformations having been obtained
through observations of fundamental standard fields), allowing us to
include these latter frames in the overall photometric solution.  Adding
images taken under non-photometric conditions does nothing to 
improve the
accuracy of our photometric transformations to the fundamental standard
system in absolute terms; however, it does allow us to increase the
precision of the photometry of individual stars relative to the mean
system defined by our 649 local photometric standards.

In our experience, photometry from datasets such as those employed here
typically display an external accuracy of roughly 0.02$\,$mag per
observation; this level of observation-to-observation scatter is 
probably
dominated by temporal and spatial fluctuations in the instantaneous
atmospheric extinction, and probably also by the difficulty of obtaining
truly appropriate flat-field corrections in the presence of such effects
as scattered light, ghosts, fringing
and spectral mismatch between the flat-field illumination and the
astronomical scene.  In the present instance, our
absolute photometry for the 649 local standards is based upon a median
number of 10, 29, and 19 observations in the $B$, $V$, and $I$
bandpasses.  Taking 0.02$\,$mag per observation as a typical external
uncertainty, we therefore expect that our photometry is referred to the
fundamental system (i.e., that of Landolt 1992) with an
absolute accuracy of order 0.006, 0.004, and 0.005$\,$mag, standard 
error
of the mean, in $B$, $V$, and $I$.  {\it Relative\/} magnitude and color
differences among different stars in our sample can be smaller than 
this,
depending upon their brightness and crowding and the number of
observations that went into each determination.

The absolute astrometry of our catalog is based upon the United States
Naval Observatory Guide Star Catalogue~I (A~V2.0; henceforth USNOGSC,
Monet \etal\ 1998),
access to which is obtained by PBS
through the services of the Canadian Astronomy
Data Centre.  The authors of the USNOGSC claim a typical 
astrometric error
of 0.25$\,$arcsec, which they believe is dominated by systematic 
errors in
the calibration procedure.  When transforming relative $(x,y)$ positions
from large-format CCD images to absolute right ascensions and 
declinations
from the USNOGSC for stars in common, we typically find root-mean-square
differences of 0.3 to 0.4$\,$arcsec in each coordinate.  Some of this is
clearly due to proper-motion displacements accumulated during the
forty-plus years between the obtaining of the first
Palomar Observatory Sky
Survey and our CCD data.  However, a significant part of the differences
is also due to the lower angular resolution of the Schmidt plates as
scanned by the Precision Measuring Machine (built and operated by the
U.~S.~Naval Observatory Flagstaff):  particularly in crowded fields 
such as
the outskirts of globular clusters, a single entry in the USNOGSC is
occasionally found to correspond to the photocenter of a close pair or a
clump of several stars in the CCD imagery.

As a result of these non-Gaussian errors (i.e., proper motions and
blending), we perform our astrometric transformations using an iterative
procedure wherein 20-constant cubic polynomials are used to approximate
the transformation of the $(x,y)$ positions measured in our CCD 
images to
standard coordinates obtained from a gnomonic projection of the right
ascensions and declinations listed in the USNOGSC.  After each iterative
step, stars lying within a certain radial distance of their predicted
positions are used to obtain an improved geometric transformation with
individual weights gradually tapering from 1.0 for perfect positional
agreement to 0.0 for stars lying exactly one critical radius from their
predicted positions; stars lying farther than one critical radius from
their predicted positions are ignored altogether.  The critical radius
distinguishing acceptable cross-identifications starts out fairly 
generous
($\sim5\,$arcsec), but then this radius is gradually reduced until it is
down to a value of one arcsecond; the iterative process then continues,
maintaining this critical radius of one arcsecond, until a stable 
list of
cross-identifications and transformation constants is achieved.  As an
indication of the percentage of entries in the USNOGSC whose 
positions we
consider no longer strictly valid, due either to proper motions or to
blending, we find that 4,709 USNOGSC entries lie within 5$\,$arcsec 
of one
of our CCD detections; 4,652 lie within 2$\,$arcsec of a CCD detection;
and 3,886 lie within 1$\,$arcsec.  When the sample is restricted to 
these
3,886 cross-identifications, the unweighted root-mean-square positional
difference is approximately 0.34$\,$arcsec in both $x$ and $y$.  Thus,
what is essentially a 3-$\sigma$ clip results in a rejection of 
some 17\%
of the ``possible'' cross-identifications that agree to within
5$\,$arcsec.

Of these 3,886 acceptable cross-identifications, only a dozen or so lie
within 200$\,$arcsec of the cluster center; none at all lie within
160$\,$arcsec.  Therefore it is possible that interpolation of the
astrometric transformations across this gap could be subject to small
systematic errors, probably not as large as 0.1$\,$arcsec.  
Throughout the
region of our field that is well populated by USNOGSC stars (including
essentially all of the stars in our present spectroscopic sample), 
we expect
systematic errors of our right ascensions and declinations on the system
of the USNOGSC to be of order 0.34$\,$arcsec/$\sqrt{3886-20}$,
or $\sim0.01\,$arcsec.  We have no independent information on the 
accuracy
with which the USNOGSC coordinate system is referred to an absolute
inertial reference frame of right ascension and declination.  Individual
{\it random\/} errors in our coordinate measurements are 
probably not much
better than 0.02$\,$arcsec on a star-by-star basis, the errors becoming
somewhat worse than this for the fainter and more crowded stars in our
photometric/astrometric sample.

Stars are identified in this paper by a name derived from their J2000
coordinates, so that star C12345\_5432 has coordinates
15 12 34.5~~+2 54 32.

\section{Spectroscopic Observations \label{section_spec}}

The initial sample of stars consisted of those
from the photometric database located more than
150 arcsec from the center of M5 (to avoid crowding) with
$16.9<V<17.35$ and with $0.86 < (V-I) < 0.96$,
i.e. subgiants at the base of the RGB. (A preliminary
version of the photometric catalog described in \S\ref{section_phot}
was used for this purpose.)
From this list, two slitmasks containing about 25 slitlets
each were designed using JGC's software.  The center of the first
field was roughly 3.2 arcmin E and 0.5 arcmin S of the center
of M5, while the center of the second field was located roughly
2.5 arcmin W and 0.5 arcmin S of the cluster center.

These slitmasks were used with LRIS (\cite{lris_ref})
at the Keck Observatory
in May 2001; three 800 sec exposures were obtained
with each slitmask.  The exposures were dithered by moving
the stars along the length of the slitlets by 2 arcsec 
between each exposure.  Because of the crowded fields,
there were often more than one suitably bright object in each
slitlet.  Hence subtraction
of sequential exposures was not possible, and they were reduced
individually using Figaro (Shortridge 1988) scripts,
then the 1D spectra for each object were summed. 
The width of the slitlets was 0.7 arcsec,
narrower than normal to enhance the spectral resolution.
LRIS-B was used with a 600 line grism giving a dispersion of
1.0\,\AA/pixel (3.0\,\AA\ resolution for a 0.7 arcsec wide slit).
This gave good coverage of the region from 3600 to 4800\,\AA,
including the key CN band at 3885\,\AA\ and the G band of CH
at 4300\,\AA.  
Technical problems related to the layout
of the slitmasks led to some of the spectra not extending
blue enough to reach the uv CN band.
The red side of LRIS was configured to use a 1200 g/mm
grating centered at H$\alpha$ with the intention of providing
higher accuracy radial velocities.  The dispersion is then
0.64\,\AA/pixel (29 \kms/pixel) or 1.9\,\AA/spectral resolution element.
Unfortunately, the LRIS-R shutter was not functioning on the first 
night,
and so only one of the two slitmasks
used has matching H$\alpha$ spectroscopy.

In addition to the primary sample described above, since
these fields are rather crowded, other stars sometimes serendipitously
fell into the slitlets, and their spectra were also reduced.
We refer to the latter as the secondary
sample.  As might be expected from the luminosity
function, most of the secondary sample consists of stars at
or just below  the
main sequence turnoff.  Two stars in the secondary sample
(C18386\_0713  and C18465\_0214)
are early type stars.  The former is almost certainly
a BHB in M5, while the status of the latter is unclear; we 
subsequently ignore both of them.

\subsection{Membership}

Given the high galactic latitude of M5 ($b=46.8^{\circ}$) and the proximity
of our fields to the center of the cluster, we expect minimal field
star contamination among the primary sample of stars.
However, the secondary sample
selection is based solely on spatial position and hence may contain
a higher fraction of field stars.
There are three indicators we use to establish membership.  The first
is deviation from the cluster sequences in a color-magnitude diagram.
Stars which are off the M5 color locus in the
($V,I$) CMD by more than 0.1 mag were considered possible non-members.
The second
is the strength of the absorption features in the LRIS multi-slit 
spectra,
ignoring the molecular bands.  The metallicity of M5 is 
sufficiently low that
near solar metallicity field stars of similar colors are easily
distinguished from cluster members.  We also
examine the radial velocity by cross correlating the red spectra
over the regime from 6400 to 6620\,\AA.

\begin{figure}
\epsscale{0.7}
\plotone{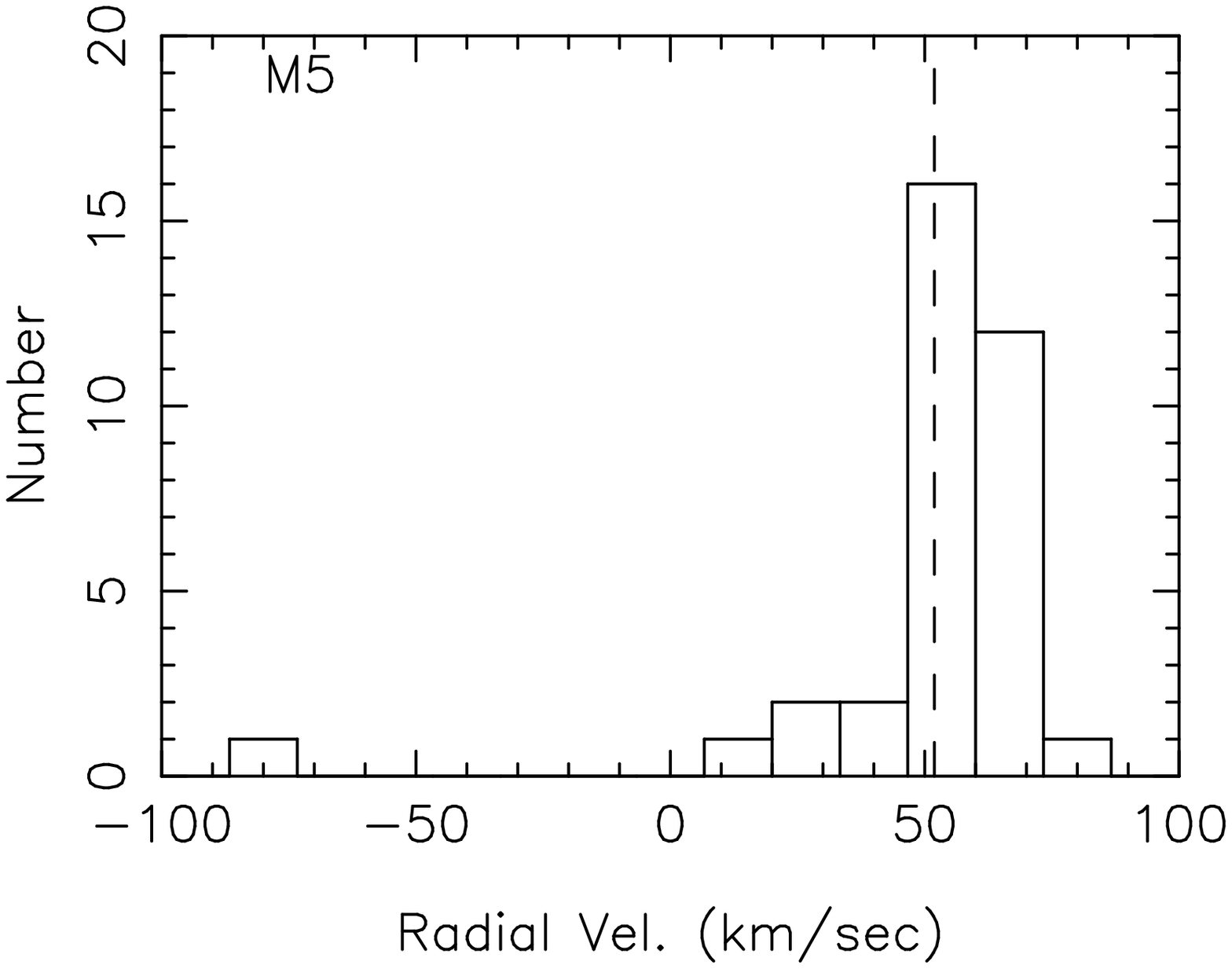}
\caption[vr_hist_ver3.ps]{The histogram of the measured heliocentric
radial velocities
for approximately half of the sample stars in M5 is shown.  The
cluster $v_r$ of Pryor \& Meylan (1993) (+51.9 \kms) is
indicated as a dashed vertical line.
\label{fig_rvhist}}
\end{figure}

The radial velocity of M5 from the compilation of Pryor \& Meylan (1993)
is +53.1 \kms, with $\sigma=4.9$ \kms.
The histogram of heliocentric radial velocities we have
measured  for 37 stars is given in Figure~\ref{fig_rvhist}.
Only one star has a radial velocity
inconsistent with membership.  Rejecting this very discrepant
object, with 36 remaining stars, we find
a mean $v_r = +52.8$ \kms, with $\sigma=11.8$ \kms.  This value
of $\sigma$ corresponds, after removing the intrinsic
stellar velocity dispersion, to an instrumental error of 0.4 pixels
in the focal plane CCD detector of LRIS-R, which seems
reasonable.

Table~\ref{table_nonmem} gives a listing of those stars in our sample
which are probably not members
based on  these three tests. The final column gives
the distance in ($V-I$) color at the fixed $V$ mag of the star
between the color of the star and that of the
cluster locus. There are two definite non-members
which appear to be more metal rich field
stars.  There are three stars which are off the cluster ($V,I$) 
color locus
by slightly more than the adopted cutoff of
0.1 mag, but which we believe to be cluster members;
all of them are crowded with a star of comparable brightness within
2 arcsec, and all are from the primary sample.
There is also one faint star from the secondary sample
which is either a blue straggler in M5 or
an early type background object, and which we subsequently ignore.

Spectra taken with the blue-side of LRIS are available for
all the stars, but at a lower dispersion.  After excluding
the probable non-members listed in Table~\ref{table_nonmem}, 
cross correlations
over the range 3800 - 4400\,\AA\ show the same 1$\sigma$ rms
dispersion as does the red side, i.e. 0.4 pixels, here equivalent to
29 \kms.  While the accuracy of these radial velocities, given
the low spectral resolution, is not sufficient to confirm
membership for any particular star, the small dispersion does 
demonstrate that almost all of the sample studied here must be
members of M5.

\subsection{The Sample of Probable Members of M5}

\begin{figure}
\epsscale{0.8}
\plotone{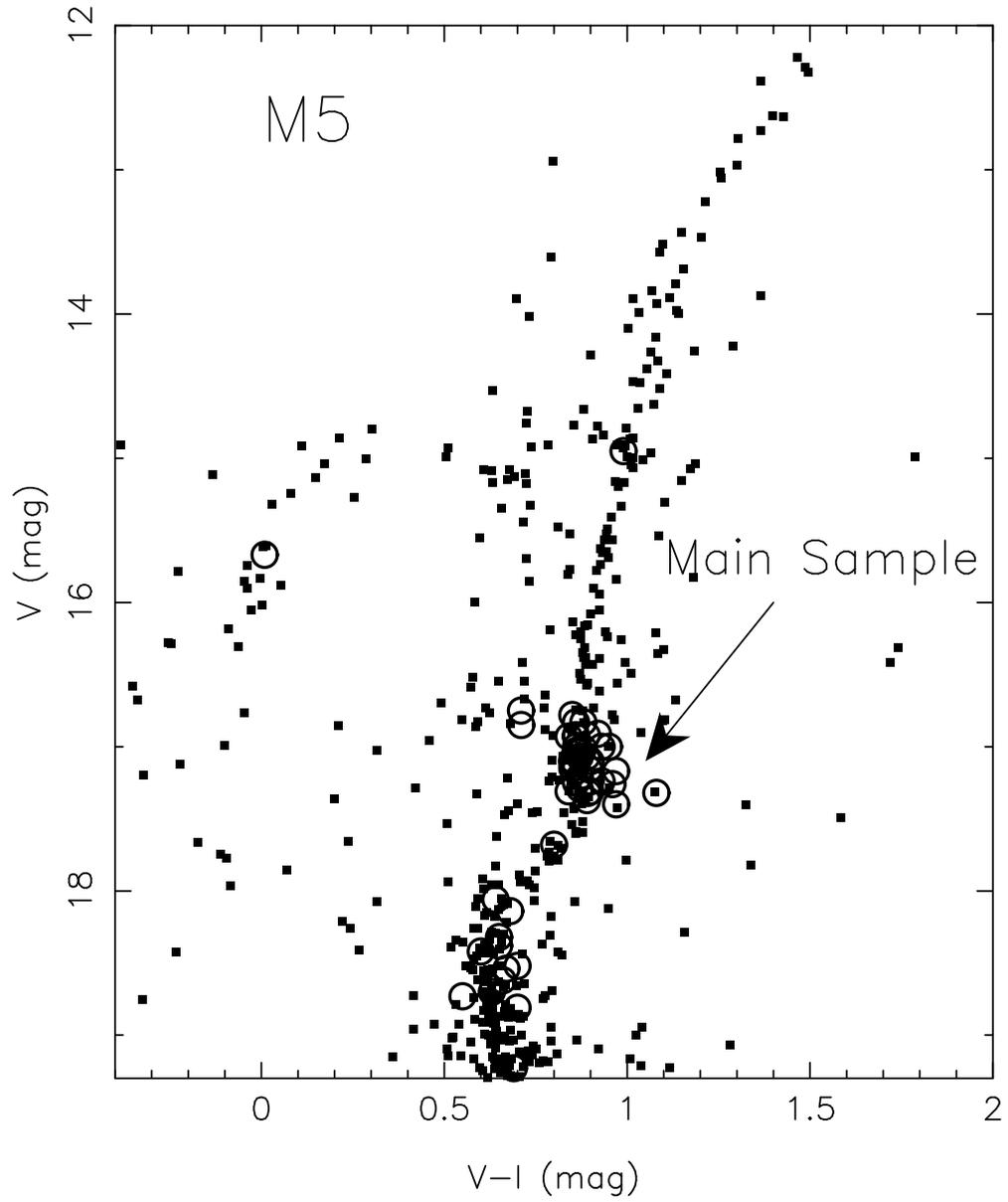}
\caption[m5_stetsonrev2_lris.ps]{The primary sample of
probable members of M5 at the
base of the red giant branch used here is shown together with
the secondary sample superposed on the $(V,I)$ CMD diagram.
\label{fig_cmdsample}}
\end{figure}

There are 59 stars left in our M5 sample after the probable
non-members and early type stars have been eliminated.  In
Figure~\ref{fig_cmdsample} these are shown superimposed on
the $(V,I)$ CMD diagram, where a decreasing fraction of stars
in the photometric database are plotted as $V$ becomes fainter to
maintain clarity in the figure. All but one of these are at least
1.5 mag fainter in $V$ than the HB in M5; there is no possible confusion
with AGB stars at these low luminosities.

\begin{figure}
\epsscale{0.7}
\plotone{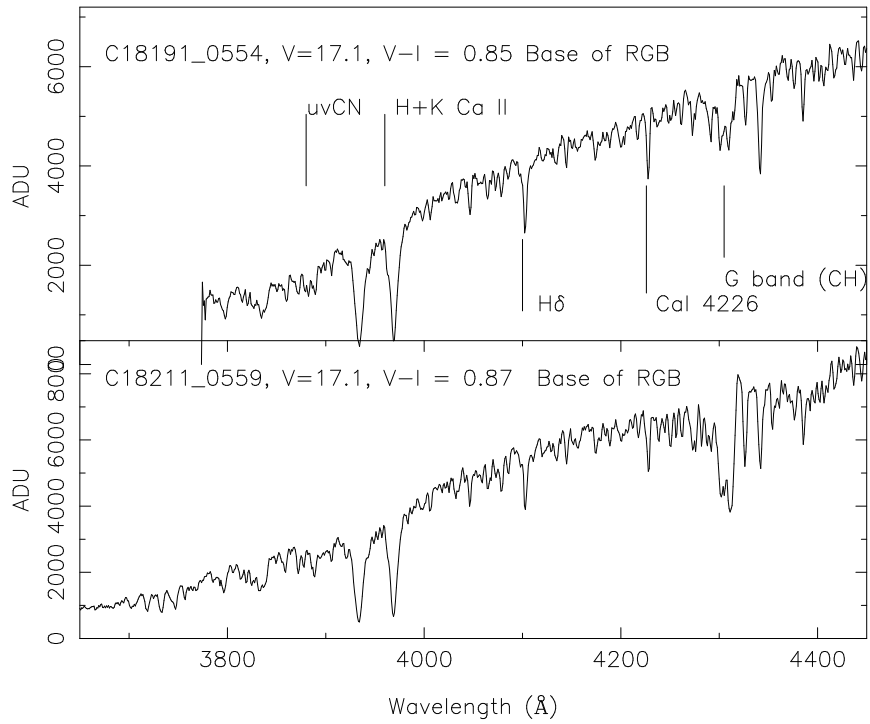}
\caption[jcohen_fig3.ps]{Part of the LRIS-B spectra of two
probable members of M5.
The stars are essentially identical in $V$ mag and $V-I$ colors and
are both located at the base of the subgiant branch.
Note the difference in the strength of the G band of CH in these two
spectra.
\label{fig_2spec}}
\end{figure}

Figure~\ref{fig_2spec} shows the region of the 4300\,\AA\ G band of CH
in the spectra of two of the stars in the primary sample.  These
stars have essentially the same stellar parameters (\teff\ and \grav)
lying at about the same place in the cluster CMD, yet their
G bands differ strongly.  The CN band near 4200\,\AA\ is too weak
to be used in these metal poor stars; the ultraviolet
CN band near 3885\,\AA\ must be used instead.
 From this figure alone, we can anticipate one of the key results
of our work, the large scatter in C abundance we will find among
M5 members at the base of the RGB at $V \sim 17.2$ mag.
We can also see that
the CN bands are much weaker, and a careful measurement will
be required, which is complicated by the difficulty of measuring
anything resembling a continuum at or near 3885\,\AA.

Table~\ref{table_photmem} lists the photometry for the
sample of 59 stars (plus one BHB star) which we believe are
members of M5.

\section{Measurement of CH and CN Indices \label{section_indices}}

\begin{figure}
\epsscale{1.0}
\plotone{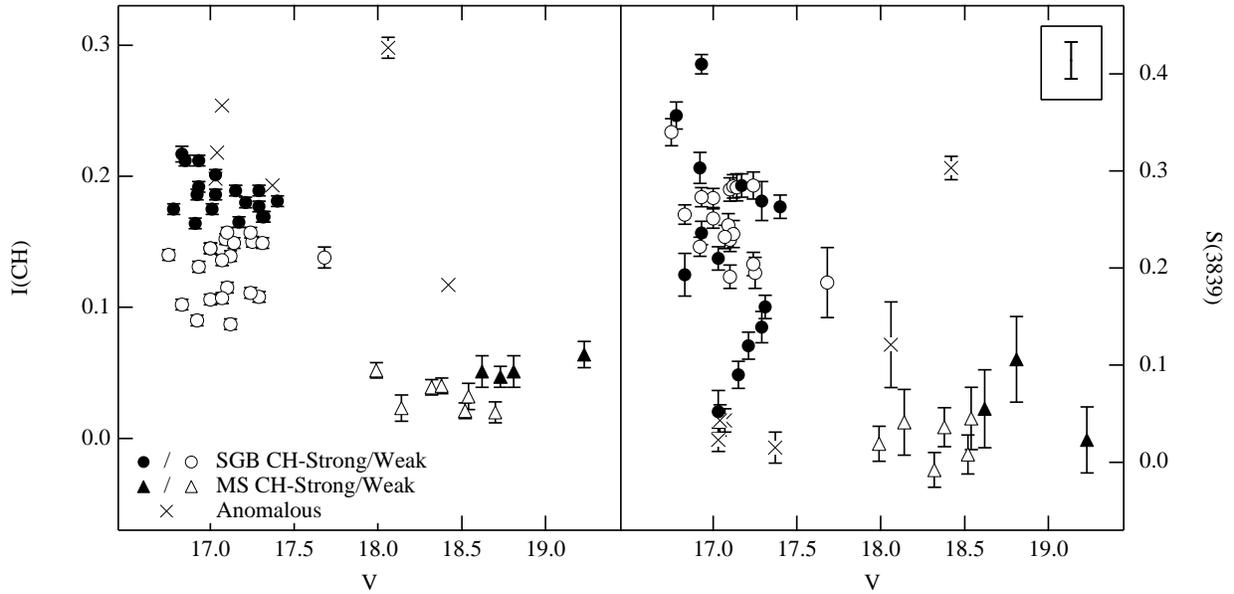}
\caption[jcohen_fig4.eps]{
The measured I(CH) and S(3839) indices are plotted for the program 
stars as
a function of $V$. The sample has been arbitrarily divided into two 
groups:
CH-strong (filled markers) and CH-weak (open). Large and 
significant star-to-star
differences exist in both CH and CN band strengths among the SGB stars.
The decreasing spread in indices with luminosity is the result of 
increasing
temperatures near the main sequence turn off.
The error bar shown in the upper-right box represents the uncertainty in
determining the slope of the uv continuum as described in
\S\ref{section_indices}.
\label{fig_chcn}}
\end{figure}

For each spectrum, S(3839) and I(CH) indices sensitive to absorption by
the 3885\,\AA\ CN band and the 4300\,\AA\ CH
band respectively (see Briley \& Cohen 2001)
were measured.
These indices are listed in Table \ref{table_obs_inds} and plotted 
in Figure \ref{fig_chcn}.
The error bars (drawn at 2$\sigma$) have been calculated strictly 
from Poisson
statistics based on the signal present in the feature and continuum 
bandpasses.

It is extremely difficult to flux spectra taken through
slitmasks because of the varying slit losses and the possibility
of atmospheric dispersion affecting the spectra.  The latter issue is
of somewhat more concern than usual to us since we are working in the
blue and UV with a narrow (0.7 arcsec wide) slit.  Carrying
out the observations with the length of the slit set to
the parallactic angle, which is the usual method for
eliminating atmospheric dispersion for single slit observations,
cannot be used for multislit observations as the position
angle is fixed by the design of the slitmask.  It is for these
reasons that no attempt was made to flux the spectra.

The reduction process for the G band index of CH,
whose feature bandpass is 4285 - 4315\,\AA~ (adjusted by 0.76\,\AA~ 
for the average
radial velocities of the present stars),
is independent of this suite of issues
as a continuum can be established both on the blue and on the
red side of the feature.  However, it is extremely difficult
to establish a continuum on the blue side of the 3885\,\AA\ CN band, and
hence we are forced to rely on a single sided index, with a continuum
bandpass determined only redward of the feature of interest.

The instrumental signature present in the raw S(3839) indices,
whose feature bandpass is 3846 - 3883\,\AA~ (shifted by 0.68\,\AA),
must be removed prior to a comparison with those predicted by
models.
This was carried out following Briley \& Cohen (2001) by fitting 
the continua
of the observed spectra with cubic splines within IRAF's ``splot'' 
facility
over the range 3600 - 5200\,\AA.
The  I(CH) indices determined from these continuum fits
(with no CH absorption included) were effectively zero
($-$0.008$\pm$0.007) as is expected from a two-sided index.

For S(3839), indices
were computed from these continua fits and the average 
(0.156$\pm$0.019)
used as a zero-point offset (as follows from the logarithmic nature 
of S(3839)) in our
comparisons. Stars with spectral coverage ending redward of 3820\,\AA\
were not included.
The process used to determine the zero point offset
for the uvCN indices thus assumes that the
transmission  $T(\lambda)$ of the atmosphere, telescope and instrument
(including slit losses) normalized to the transmission at some 
fiducial
wavelength, e.g. 4000\,\AA, is a function of wavelength which does 
not change
among the spectra of the M5 stars over the relevant wavelength
range, i.e. that of the two bands used to define S(3839), 
from 3840 to 4000\,\AA;
this is a less stringent requirement than would be needed to flux
the spectra.

Slit alignment errors which vary from star to star (possible with
the narrow 0.7 arcsec wide slits used here) combined with
atmospheric dispersion are the most likely way to
introduce a variation in the normalized $T(\lambda)$
from star to star, which would in turn introduce a dispersion in
S(3839) through variation of the zero point from that adopted above.  
The two slitmasks
used for our M71 sample were
observed at mean airmasses of 1.08 and 1.25 respectively.
This should be adequate to avoid serious problems with
atmospheric dispersion.  Cohen \& Cromer (1988)
discuss this issue in more detail.

We have tested the correlation between fit uv continuum slope
and various parameters, including
RA, Dec, luminosity and S(3839).  The correlations range from
not statistically significant to marginally significant, with
the correlation between uv continuum slope and S(3839) being
the most significant (correlation coefficient of 0.48)
and that with luminosity being the least significant (correlation
coefficient 0.09).

The small corrections derived from these correlations
have a maximum value of about 0.015
and a minimum value of 0.000, which is
less than 5\% of the total range and
smaller than the quoted uncertainties.
Given that the S(3839) variations among the stars in our sample
of subgiants in M5 cover a range of 
roughly 0.3 in S(3839), these small corrections,  if implemented, 
would have little affect on our resulting C and N abundances, 
and we have elected not to include them.

As can be seen in Figure \ref{fig_chcn}, a substantial star-to-star
variation is seen among I(CH) indices for the stars in our M5 sample,
even if one considers only subgiant branch (SGB) stars in 
similar evolutionary states.
To aid in interpretation, the stars in Figure \ref{fig_chcn} have 
been arbitrarily divided
into CH-weak (open symbols) and strong groups (filled symbols).

Six of the stars with the most extreme CH or CN indices are
denoted as anomalous.  Of these, five have radial velocities,
all consistent with membership in M5.
Four of these stars
are in the sample at the base of the RGB and show
very strong I(CH) with very weak CN.  C18225\_0537 
has $V = 18.06$ and shows an enormous I(CH) with a strong  S(3839) 
as well.  C18206\_0533 ($V = 18.42$) shows
modest I(CH) but very strong uv CN.  For the brighter of these
stars, the anomaly noted in the final summed spectra
is clearly present on each individual spectrum of the object,
and hence it is highly unlikely that these anomalies are due
to CCD defects or other such problems.
We plan to check the membership
of each of these stars with higher dispersion spectra as soon
as possible, but at the present time, we believe that most, if not
all, of these anomalous stars are in fact members of M5.

\begin{figure}
\epsscale{0.7}
\plotone{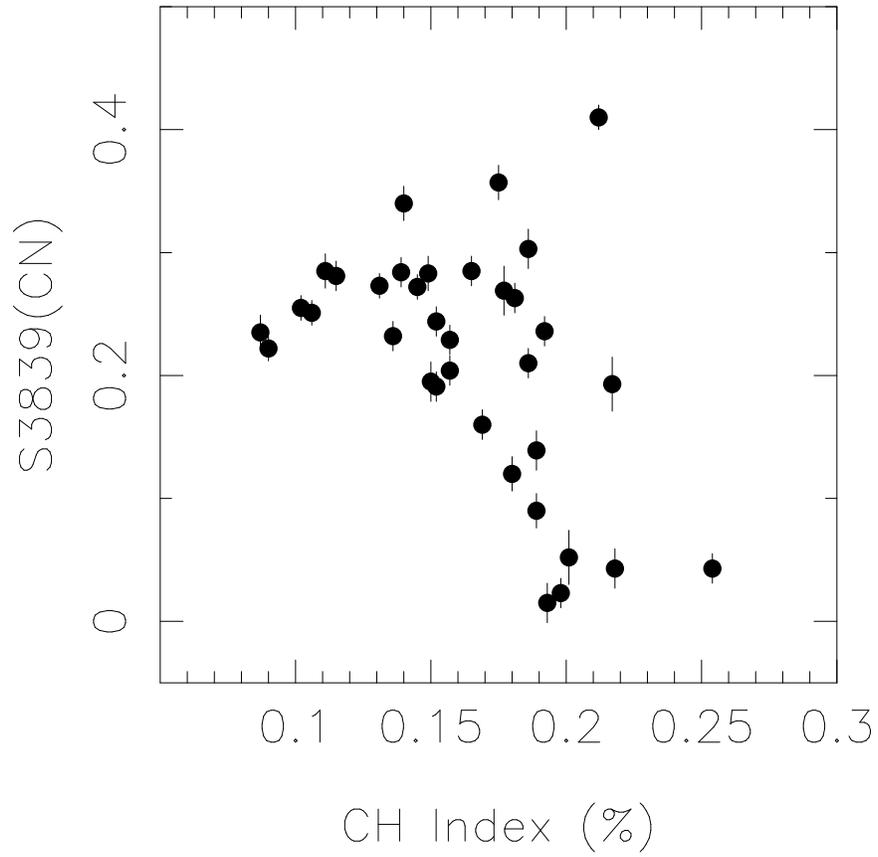}
\caption[jcohen_fig5.ps]{
The CH and uvCN band indices are plotted against each other for
the 43 stars with $16.5 < V < 17.5$ mag at the base of the RGB in M5.
The two stars with the strongest CH indices were classified
as anomalous.
\label{fig_rawchcn}}
\end{figure}

We proceed by restricting our attention to the subgiants at the base of
the RGB in M5, specifically to those with $16.5 < V < 17.5$ mag.
Figure~\ref{fig_rawchcn} displays I(CH) versus S(3839) for this
group of 43 stars.  Again the very large range in each index within
this small range in evolutionary state, hence small range
in \teff\ and \grav, is apparent.  Also apparent is a strong
anti-correlation between the strength of the CH band and that of
the CN feature.  The correlation is not perfect, and tends to
turn over among the stars with the weakest I(CH).
The bimodal distribution of CN band strengths found among the
most luminous M5 giants
by Smith \& Norris (1983) is also not apparent
in the present sample.
This may be the result of the
higher temperatures
(and correspondingly weaker CN bands) of our stars, combined with a 
finite
observational error tending to blur out the bimodality.

The large range in C abundances which we suspect to be present in the
M5 subgiant sample creates an unusual situation with regard to the
expected strength of the CN features.  Normally, since there is
more carbon than nitrogen, the N abundance controls the amount
of CN.  However, if C is highly depleted, there can be fewer carbon
atoms per unit volume than nitrogen atoms, and C will control the
formation of CN, as suggested by Langer (1985).  
Within the small
range of \teff\ covered by the M5 subgiant sample, we might expect
the observed relationship between the molecular band indices
of CH and of CN to be non-monotonic, as is in fact shown in
Figure~\ref{fig_rawchcn}.  In particular, in this figure, the
stars with the weakest CH do not have the strongest CN indices.

\subsection{The Main Sequence Turn Off Stars}

\begin{figure}
\epsscale{1.0}
\plotone{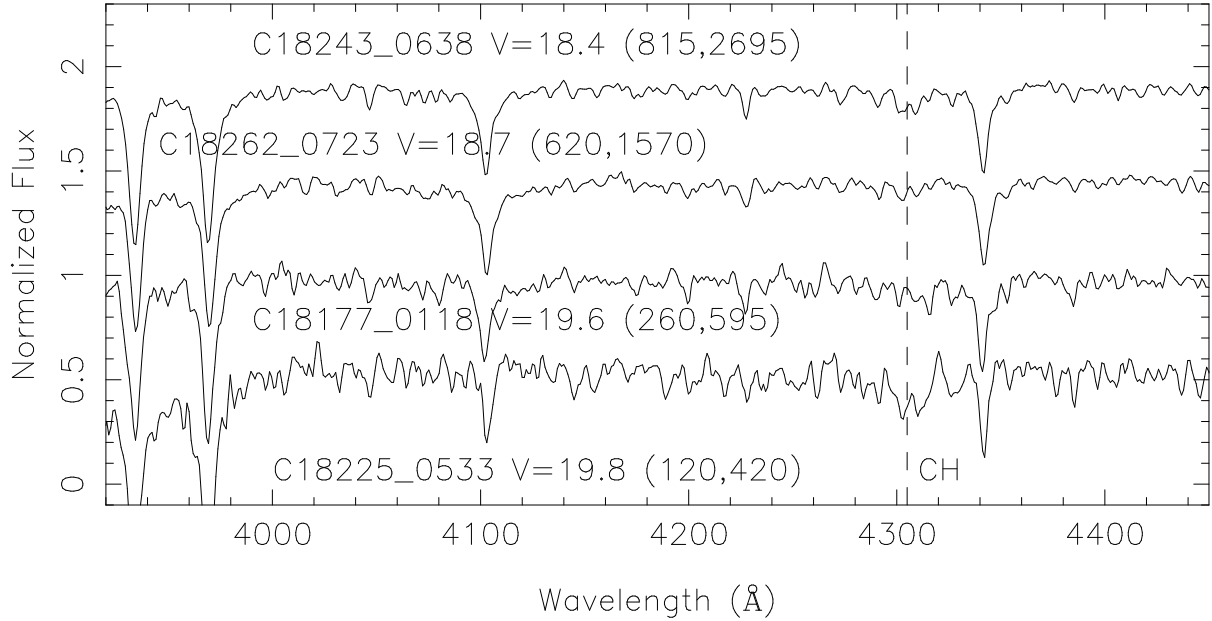}
\caption[jcohen_fig6.ps]{The region of the G band of CH in four main
sequence stars in M5 going from top to bottom in order of increasing
$V$ mag.  A constant arbitrary vertical offset
between the spectra has been imposed for clarity.
The label for each object gives in parentheses the DN/pixel at
the blue and red end respectively in the original spectrum so that
the signal-to-noise ratio can be calculated.
The spectra have been slightly smoothed, then a continuum was fit
from 3900 to 4500\,\AA, and the spectra were normalized.  In addition
to looking at the strength of the CH band, note that
the strongest atomic features,
particularly  the Ca I lines at 4226\AA\ and the
Fe I lines at 4046 and 4384\,\AA, are also becoming stronger
as \teff\ decreases.
\label{fig_msch}}
\end{figure}

There are twelve stars in our sample with $V > 18.2$ which are believed
to be members of M5 at or slightly fainter than the
main sequence turnoff (MSTO).  Their molecular bands are, as expected,
much weaker due to their higher \teff.  However,
it does appear that variations in I(CH) are still present, although
not as large in extent as are seen among the subgiants.
Figure~\ref{fig_msch} presents the
region of the G band of CH in four of the main sequence stars
in M5 in order of increasing $V$ mag.  They
span a total range of 1.4 mag in $V$ from the top of the
main sequence downward in luminosity and cooler in \teff.
The spectra have been slightly smoothed using a Gaussian
with FWHM of 5 pixels, then a continuum was fit
from 3900 to 4500\,\AA, and the spectra were normalized.  The
signal-to-noise ratio of the spectra decreases as the magnitude 
increases.
The CH band is marginally stronger in the first star than in the second,
and strongest in the coolest star, much stronger than in
a main sequence star only 0.2 mag brighter. The strongest atomic 
features,
particularly  the Ca I lines at 4226\,\AA\ and the
Fe I lines at 4046 and 4384\,\AA, are also becoming stronger
as \teff\ decreases.
Higher precision data and a larger
sample of main sequence turnoff stars will be required to determine
in a definitive way the trends among CH and CN.

\section{Comparisons with Synthetic Spectra \label{section_cnabund}}

Clearly the pattern of abundances underlying the CH and CN band
indices of Figure  \ref{fig_chcn} cannot be interpreted on the
basis of band strengths alone - we must turn to models.
The technique employed is similar to that of Briley \& Cohen 
(2001), where the
region of the CMD of interest is fit by a series of models whose 
parameters are taken
from the Bergbusch \& VandenBerg (1992) O-enhanced isochrones.
For M5, the 16 Gyr isochrone with
[Fe/H] = --1.26  was used to select a set of representative model points
which are listed in Table \ref{table_models}.  Model stellar
atmospheres were then generated using
the Marcs model atmosphere program (Gustafsson \etal\ 1975) at the
\teff, \grav\ of these points.
 From each model, synthetic spectra were calculated using the SSG 
program (Bell \&
Gustafsson 1978; Gustafsson \& Bell 1979; Bell \& Gustafsson 1989; 
Bell, Paltoglou,
\& Tripicco 1994) and the line list of Tripicco \& Bell (1995).
Each synthetic spectrum was initially computed from 3,000 to 
12,000\,\AA\ in
0.1\,\AA\ intervals and convolved with $V$ and $I$ filter curves as
described in Gustafsson \& Bell (1979) and Bell \& Gustafsson (1989) to
yield colors appropriate to each model.

\begin{figure}
\epsscale{0.7}
\plotone{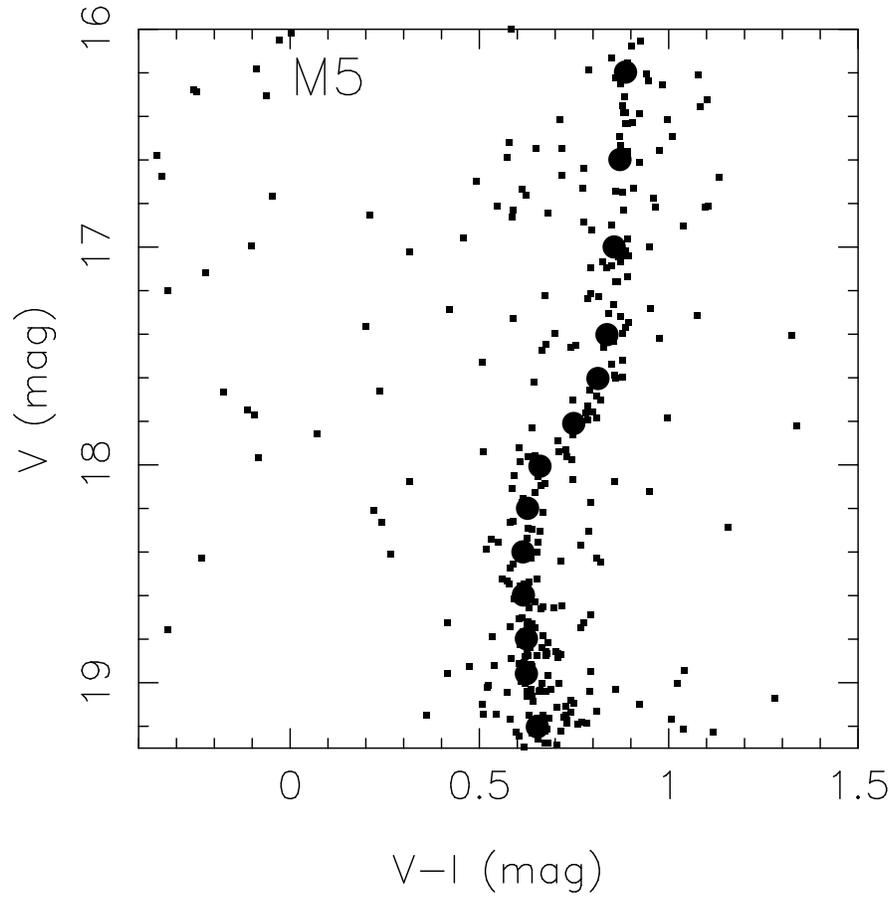}
\caption[cmd_briley_model2.ps]{The model points for the 
grid of stellar atmospheres and spectral syntheses
obtained from the isochrone are shown 
superposed on the $(V,I)$ CMD diagram of M5 in the region of the
subgiant branch and the upper main sequence.
\label{fig_brileycmd}}
\end{figure}

Figure~\ref{fig_brileycmd} shows these model points
superposed on the M5 $V,I$ CMD.
We assume stellar masses as given by the isochrone,  a microturbulent
velocity of 2 km/s, and the cluster parameters given in \S\ref{intro}.
As is apparent from the Figure, the isochrone 
closely follows the M5 CMD.  We have compared \teff\ derived 
from the isochrone itself,
which values we subsequently adopt, with \teff\ derived from the
observed $V-I$ colors of each of the subgiants
individually.  We use the grid
of predicted colors from Houdashelt, Bell \& Sweigart (2000)
for this purpose, together with the \grav\ from the isochrone.
(The dependence on \grav\ is relatively small.)
The mean difference is $\sim+70$K, with a large $\sigma$ of
90 K.  While some part of this mean difference might be photometric errors,
if interpreted as an error in matching 
the color of the isochrone at a fixed $V$ with 
the locus of the M5 stars, this corresponds to a systematic difference
in $V-I$ of about $0.03$ mag.  A hint of this is seen 
in Figure~\ref{fig_brileycmd},
where the cluster locus does appear slightly redder than the
adopted model points near $V {\sim}$17.5. Ignoring these
small differences, we 
therefore
believe that assigning \teff,\grav\ utilizing the $V$ mag plus
the isochrone is a reasonable way to represent the mean properties
of the M5 subgiant sample as a function of luminosity.

Using these models we have further calculated a series of
synthetic spectra at higher
resolution (0.05\,\AA\ intervals) from 3500 to 5500\,\AA\ with a
variety of C, N, and O
abundances.
These spectra were then smoothed to the resolution of the observed 
spectra and
the corresponding I(CH) and S(3839) indices measured.
Zero point shifts were also measured from the synthetic spectra by 
loading them into
IRAF and performing the continuum fits discussed above for the 
observed spectra.
The resulting offset for S(3839) was 0.046$\pm$0.010 and 
0.003$\pm$0.001 for I(CH).

\begin{figure}
\epsscale{0.7}
\plotone{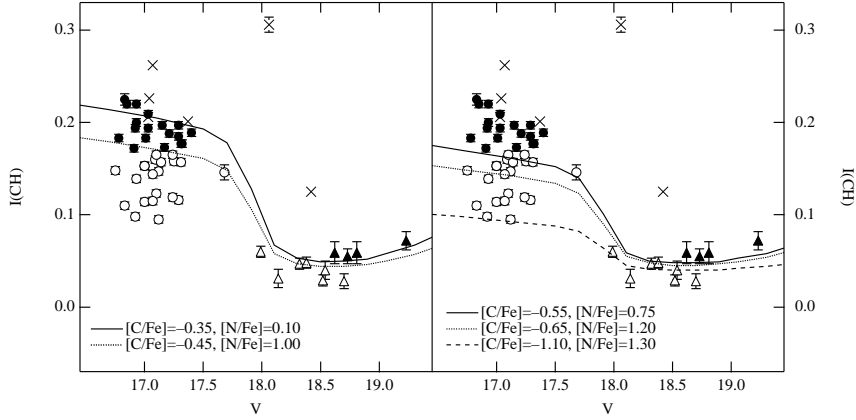}
\caption[jcohen_fig8.eps]{
The observed I(CH) indices are plotted with calculated band 
strengths for
several differing C (and N) abundances as listed in Table 
\ref{table_models}
as a function of $V$ magnitude.
As in Figure \ref{fig_chcn}, the stars have been divided into two 
groups based
on CH-band strength.
Note that the range of CN-band strengths observed requires nearly a 
0.75 dex
star-to-star variation in [C/Fe] among the SGB stars.
\label{fig_ssg_ch}}
\end{figure}

The values of I(CH) from five sets of models with [C/Fe] from 
--0.35 to --1.10,
[O/Fe] = +0.25, and \ciso = 10 are plotted with the observed indices in
Figure \ref{fig_ssg_ch}.
The spread in [C/Fe] among the M5 SGB stars appears well 
represented by this
range, which is very similar to that reported among a sample of 
more luminous
M5 SGB stars by Briley \etal\ (1992) (although a lower value of
[Fe/H] = --1.40 was chosen in their study).
We note the very low C abundances implied by the weakest I(CH) 
indices among
the SGB sample, which are consistent with the low C abundances 
observed among
the brightest M5 giants (Smith \etal\ 1997).
However, in the albeit small Smith \etal\ sample of
five giants, no analog to the
CH-strong stars (at least in terms of [C/Fe]) were observed.

\begin{figure}
\epsscale{0.8}
\plotone{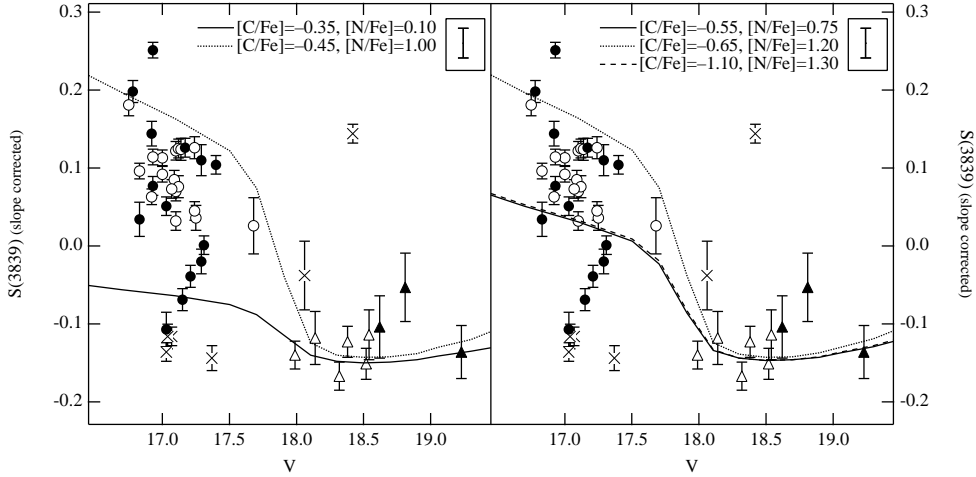}
\caption[jcohen_fig9.eps]{
Observed S(3839) indices are plotted with the model indices from 
Table \ref{table_models}
against $V$ magnitude.
As with Figure \ref {fig_ssg_ch}, a significant spread in 
abundances is present to within
0.5 mags of the MSTO.
Among the MSTO stars themselves, the higher temperatures have 
resulted in CN
band strengths too weak to  measure accurately in the present spectra.
The error bar shown in the boxes indicates the uncertainty in the 
S(3839)
offset due to the slope of the uv continuum (see 
\S\ref{section_indices}).
\label{fig_ssg_cn}}
\end{figure}

The large range in C abundances indicated in Figure 
\ref{fig_ssg_ch} leads to a
difficulty in assessing N abundances via CN band strengths, as
discussed in \S\ref{section_indices}.
Among the CH-strong SGB stars (see the left panel
of Figure \ref{fig_ssg_ch}) there is a small
($\approx$ 0.1 dex) spread in [C/Fe], with N being the minority
species controlling the formation of CN.
Thus, the CN band strengths of these stars can be assumed to 
roughly represent
the [N/Fe] abundances.
This is demonstrated in Figure \ref{fig_ssg_cn} (left), where the 
S(3839) indices are
compared to those from two series of
synthetic spectra, the first with [C/Fe] = --0.35, 
[N/Fe] = +0.10 and the second with
[C/Fe] = --0.45, [N/Fe] = +1.00 (with [O/Fe] = +0.25 and \ciso=10
for both, as in Figure \ref{fig_ssg_ch}).
Clearly the star-to-star spread in N abundances of the CH-strong 
SGB stars approaches a factor of 10. 

For the CH-weak stars, the situation is more complex - the lower C 
abundances
can regulate CN formation.
Thus, as suggested by the discussion of \S\ref{section_indices},
the large spread in [C/Fe] among the CH-weak
stars essentially destroys any relation between S(3839) and N abundance.
This is illustrated in Figure \ref{fig_ssg_cn} (right) where
S(3839) indices from model
spectra with three different combinations of [C/Fe] and [N/Fe] are 
plotted.
While [C/Fe] = --0.65, [N/Fe] = +1.20 results in the stronger of 
the CN bands,
the models with [C/Fe] = --0.55, [N/Fe] = +0.75 have CN band 
strengths essentially
indistinguishable from [C/Fe] = --1.10, [N/Fe] = +1.30.
Indeed, for the most C-poor of the current SGB stars, significant 
enhancements of
N are required to reproduce the observed S(3839) indices.

We defer an attempt to deduce C and N abundances among the
M5 main sequence turn off stars until a larger sample of spectra of 
such stars
with higher signal-to-noise ratio than the present set of spectra
becomes available.

\section{Inferred C and N Abundances Among the Subgiants 
\label{section_abund}}

\begin{figure}
\epsscale{0.8}
\plotone{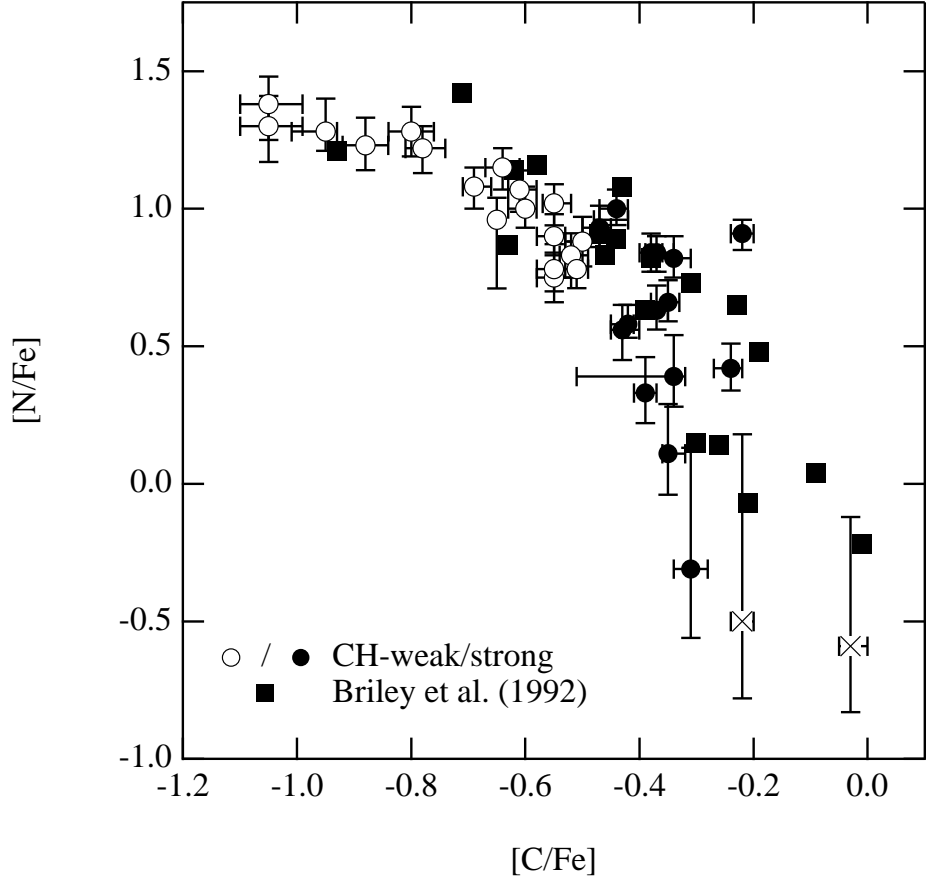}
\caption[jcohen_fig10.eps]{
The resulting [C/Fe] and [N/Fe] abundances for the M5 SGB stars in Table
\ref{table_obs_inds} are plotted.
A strong C versus N anti-correlation is evident which also compares 
well with
the relation from Briley \etal\ (1992)  among a 
sample of more luminous
cluster members (plotted as filled squares).
The presence of such an anticorrelation, although suggestive of the 
presence of
atmospheric material exposed to the CN-cycle, is difficult to
explain via internal processes given the evolutionary state of the
present sample of stars.
\label{fig_c_vs_n}}
\end{figure}

To disentangle the underlying C and N abundances from the CH and CN band
strengths, we have simultaneously fit the [C/Fe] and [N/Fe] 
abundances corresponding to
the observed I(CH) and S(3839) indices of the SGB stars.
The temperatures and gravities used for the models were chosen
based on the $V$
magnitudes of the observed stars in the present photometry as
fit to the isochrone
in Figure \ref{fig_cmdsample} using cubic splines.  The validity of this
procedure was established in \S\ref{section_cnabund}.
The abundances of C and N were then varied until the model I(CH) and
S(3839) indices matched the observed values as closely as possible, 
including
the derived zero points.
The mean rms error in the fits to the observed indices was 0.005.
For all stars, a value of [O/Fe] = +0.25, \ciso = 10, and a 
microturbulent
velocity of 2 km/s was assumed (the sensitivity to these 
assumptions will be
explored below).
The resulting C and N abundances for the M5 SGB stars are listed in 
Table
\ref{table_obs_inds} and plotted in Figure \ref{fig_c_vs_n}.

Immediately apparent in Figure \ref{fig_c_vs_n} is a dramatic 
anti-correlation between
the C and N abundances of the M5 SGB stars.
Also plotted are the [C/Fe] and [N/Fe] abundances of a
sample of more luminous M5 SGB stars reported in Briley \etal\ (1992)
(note we have plotted the abundances
from their [Fe/H] = --1.25 models), which compare very well with 
the present results.
The error bars were determined by repeating the fitting process 
while including shifts in the
observed indices of twice the average error among the SGB
indices as derived from
Poisson statistics (0.005 in I(CH) and 0.02 in S(3839)).
The shifts were included in opposing directions (e.g., +0.005 in 
I(CH) and --0.02 in S(3839),
followed by --0.005 in I(CH) and +0.02 in S(3839)) and reflect 
likely errors in the
abundances due to noise in the spectra.

\begin{figure}
\epsscale{0.9}
\plotone{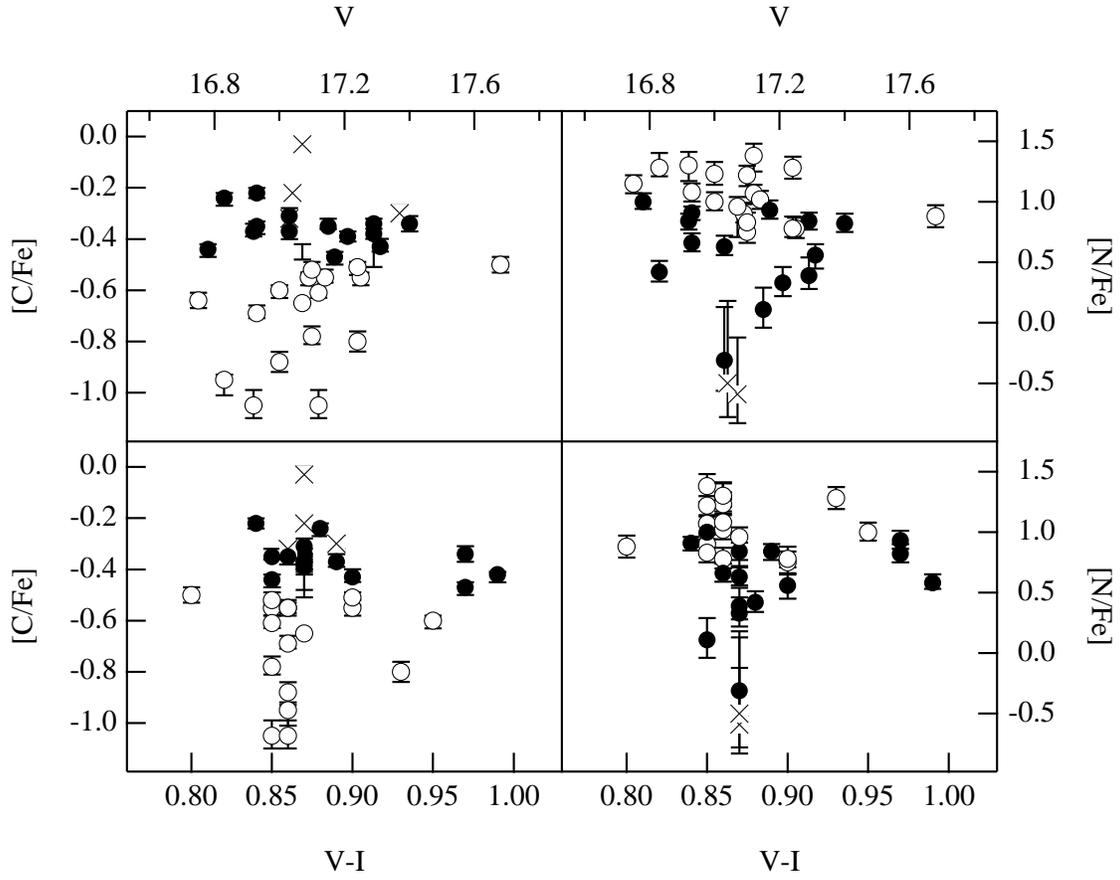}
\caption[jcohen_fig11.eps]{
The derived C and N abundances are plotted against photometry for 
the SGB stars
of Table \ref{table_obs_inds}. No systematic trends with either 
luminosity or
temperature are apparent in the abundances.
\label{fig_c_n_photo}}
\end{figure}

Of much greater concern are the systematic errors, as they have the 
potential
to exaggerate any C versus N anticorrelation (i.e., stars with 
overly low
C abundances will naturally require greater abundances of N to 
match the observed
CN band strengths).
To assess the role of many of our assumptions in shaping Figure 
\ref{fig_c_vs_n},
we have chosen three representative SGB stars and again repeated
the fitting of the CN and CH band strengths with different values 
of [Fe/H],
[O/Fe], \ciso, etc.
These results are presented in Table \ref{table_changes}, where it 
may be seen that
the sensitivity of the derived C and N abundances to the choice of 
model parameters
is remarkably small (well under 0.2 dex for reasonably chosen 
values), as would
be expected from these weak molecular features.
We have also plotted the SGB C and N abundances as both functions 
of $V$ and
$V-I$ colors in Figure \ref{fig_c_n_photo} to evaluate possible 
systematic effects
with luminosity (albeit over a small range) and 
temperature;  none appear to be present.

Evaluating the absolute abundance scale is more difficult as 
external comparisons
are limited. For the main sequence stars in 47 Tuc,
we can compare the results of Briley \etal\ (1991, 1994), carried out
in a manner fairly similar to the present work, with the independent
analysis of a different sample of stars by Cannon \etal\ (1998).
This suggests we may be systematically underestimating
the absolute C abundance by about 0.15 dex, and overestimating
the N abundance by about 0.2 dex. However, such a shift cannot
account for the large range in C and N abundances apparent in Figure
\ref{fig_c_vs_n}. We therefore conclude that the C versus N 
anti-correlation
among the SGB stars in Figure \ref{fig_c_vs_n} is indeed real.

\section{Review of Previous Studies of C and N in Globular Clusters}

\subsection{The Case of M5}

There have been several previous studies of the C and N abundances 
in M5.
Smith \& Norris (1983) measured CH and CN indices for a sample of
29 stars near the RGB tip.  They found a bimodal distribution of
CN absorption.  Langer \etal\ (1985) reobserved six stars
selected from this sample, three CN-strong and three CN-weak, and 
calculated
the change in mean C and N abundance between the two groups.
Briley \etal\ (1992) observed 14 subgiants with $16.2 < V < 16.5$,
analyzed in a manner similar to the present paper, to find
large stochastic variations from star-to-star of a magnitude similar
to those found here for still fainter subgiants,  and with a clear
detection of bimodality for the S(3839) indices of these subgiants.

A question of considerable import for understanding the behavior of
C and N in M5 is whether or not there is any change in the
mean C abundance as one moves from the RGB tip down to the base of
the subgiant branch.
Any effect dependent on evolutionary state is an argument favoring
internal mixing
rather than an external (primordial or pollution) origin
for this phenomenon.
While this issue is complicated by the need
to convert between three different definitions for the strength
of absorption at the G band of CH, the best that
can be said at the present time is that any difference in the
mean [C/H] between the tip and the base of the RGB in M5 must be less
than 0.3 dex. We seem to be in the peculiar position of having a
better sample of stars with published C and N abundances
between $16.0 < V < 17.5$ than exists at the RGB tip.
We have already initiated an effort to remedy this and to refine this
crucial number.

\subsection{Other Globular Clusters \label{section_othergc}}

Cohen (1999) has shown that the C/N anti-correlation seen among
bright RGB stars in M71
extends to the stars at the main sequence turnoff and even fainter 
in M71.
At the level of the main sequence, the distribution of
both the CH and CN indices appears to be bimodal.
Briley \& Cohen (2001) have used model atmospheres combined with
synthetic spectra in an approach similar to that of the present paper
to show that the C and N abundance range
seen at the level of the main sequence is comparable to that seen
among the bright red giants of M71 by many previous studies,
the most recent of which is Briley, Smith \& Claver (2001). To
reproduce the  M71 CH and CN bands requires a range in C
of a factor of 2 starting from [C/Fe] = 0.0 and a range in N
of a factor of 10, starting from an initial N enhancement
of a factor of 2.5 ([N/Fe] = +0.4 dex, together with a
decrease in O from [O/Fe] = +0.4 to +0.1.  These numbers
are very similar to what is required to explain the M5 subgiants,
with the exception of [C/Fe], which exhibits a larger
variation among the M5 subgiants (a factor of 6 versus 2).

Norris, Freeman \& DaCosta (1984) studied the CH and CN indices
among a sample of 112 bright RGB stars in 47 Tuc (a high metallicity
cluster, with [Fe/H] similar to that of M71), while Cannon \etal\ (1998)
observed a sample of comparable size from the base of the RGB to the
upper main sequence.  Large stochastic star-to-star variations
with similar anti-correlations
between C and N abundances and with a bimodal behavior of CN
were found at all luminosities examined.  The range of mean abundances
between the CN-strong and CN-weak groups appeared unchanged from the
lower RGB to the main sequence, with the CN-strong group having
$<$[C/H]$> = -0.15$, $<$[N/H]$> = +1.05$ dex, as compared to
$<$[C/H]$>$ = +0.06, $<$[N/H]$> = +0.20$ dex for the CN-weak 
stars.  The N
range is comparable to that found here for M5, but again, as with 
M71, the C range is a little
smaller in 47 Tuc.  Similar to the situation in M5, large 
samples exist both at the tip
of the RGB and at the main sequence in 47 Tuc, but they are not
well tied together, and it is not possible from the published
papers themselves to establish the
magnitude of any systematic change in the mean $<$[C/H]$>$ with
luminosity, beyond a statement that it cannot be large.

Determinations of the \ciso\ ratio have been made for the
brightest RGB stars in 47 Tuc by Brown, Wallerstein \& Oke (1990)
and Bell, Briley, \& Smith (1990), as well as in M71 by Briley 
\etal\ (1997).
They obtain \ciso\ $\sim$4 -- 8, correlated with CN-band strength, 
implying that
the surface C-poor, N-rich envelope material has been exposed to 
proton capture.

The seminal survey of C and N in the intermediate metallicity clusters
M3 and M13 (two clusters of slightly lower metallicity than M5)
by Suntzeff (1981) established the existence of a bimodal distribution
of CN band strength, at least for $M_V \lesssim -0.4$, with
stochastic variations in C and N abundance from star to star in each
cluster.
Smith \etal\ (1996) combine their C and N abundances for a sample
of stars near the tip of the RGB in each cluster
with published O abundances to show that the total of C+N+O is constant
for these luminous giants.

The only other cluster with similar data available for its MSTO 
stars is NGC
6752, a cluster of similar metallicity to M3 and M13.
Suntzeff \& Smith (1991) observed both CH and CN variations among its
MSTO stars. Gratton \etal\ (2001) report an anticorrelation between 
O and
Na, as well as between Mg, Al and C and N, similar to the relations
found among the luminous giants. Suntzeff \& Smith (1991) also found
low (3 - 10) \ciso\ ratios, which they attributed to some mixing 
taking place
but with primordial variations already in place.

M92 was studied in detail by Carbon \etal\ (1982), who used the NH
band at 3360\,\AA, as the CN bands become too weak at such low 
metallicities.
They studied 45 giants and subgiants with $M_V < +2$ and found
strong stochastic star-to-star variations over their full 
luminosity range
of about a factor of 3 in C and a factor of 10 in N.  However,
C and N are not in general anti-correlated.
The data for fainter stars is limited at present (at least by the
standards of the large samples we have achieved for M71 and for M5).
However, it is
clear that in addition to the variations, there is an easily
detectable systematic decrease in the mean C abundance by
about a factor of 10 from the base of the RGB to the
top of the RGB (Langer \etal\ 1986, Bellman \etal\ 2001).
They argue that this is the key
fact in establishing that internal mixing is the dominant effect
controlling the C and N abundances in M92.

M15, with metallicity similar to M92, was studied by Trefzger 
\etal\ (1983).
In a sample of 33 bright giants reaching to $M_V=+1.2$, they found
strong stochastic star-to-star variations, and found that the
mean $<$C/C$_0>$ declines with advancing evolutionary state, but
the mean N abundance does not change.  Furthermore, they found
that  N/N$_0$ was so large for about 1/3 of the sample that it
could not be explained even by converting all C into N.

CH in the upper RGB stars in NGC 6397, another metal poor cluster,
was explored by Bell, Dickens \& Gustafsson (1979).

\section{Discussion \label{section_discussion} }

The primary facts that we have established for M5 are that there are
strong stochastic variations from star-to-star
of both C and N, with C and N anti-correlated, similar to those
seen in M71.  These variations
are definitely present among the subgiants at the base of the
RGB to $M_V {\sim} +3$ and appear to extend to the main sequence
stars as well.  A question of considerable import for 
understanding the behavior of
C and N in M5 is whether or not there is any change in the
mean C abundance as one moves from the RGB tip down to the base of
the subgiant branch such as is clearly seen in M15 and M92.
As discussed in \S\ref{section_othergc} above,
the best that
can be said at the present time is that,
unlike the case of M92 or M15,
any systematic decline in the C abundance with evolutionary state
as one moves up the RGB appears to be small, $<0.3$ dex.

\subsection{Implications for Stellar Evolution}

A classical review of post-main sequence stellar evolution can be
found in Iben \& Renzini (1983).  Their description of the consequences
of the first dredge up phase, the only dredge up phase to
occur prior to the He flash,
indicates that a doubling of the surface N$^{14}$ and a 30\% reduction
in the surface C$^{12}$ can be expected, together with a
drop in the ratio of
C$^{12}$/C$^{13}$ from the solar value of 89 to $\sim$20,
as well as a drop in surface Li and B by several orders of magnitude.
Observations of field stars over a wide range of luminosities
conform fairly well to this picture, see e.g. Shetrone \etal\ (1993),
Gratton \etal\ (2001), although additional mixing of Li and
lower than predicted ratios of C$^{12}$/C$^{13}$ seem to 
occur even among field stars (do Nascimento \etal\ 2000).

To match the observations of variations in
abundances among globular cluster red giants which far exceed those
described above, additional physics must be introduced into
calculations of dredge up in old metal poor stars.
Relevant phenomena include
meridional mixing as described by Sweigart \& Mengel (1979)
as well as turbulent diffusion (see Charbonnel 1994, 1995)
and the insights of Denissenkov \& Denissenkova (1990)
concerning the importance
of the $^{22}$Ne($p,\gamma)^{23}$Na reaction as a way to produce
p-burning nuclei.  

The clear prediction of the most current calculations of this type
by Denissenkov \& Weiss (1996), Cavallo, Sweigart \&  Bell (1998)
and Weiss, Denissenkov \& Charbonnel (2000) is that
the earliest that deep mixing can
begin is at the location of the bump in the luminosity function
of the RGB which occurs when the H-burning shell crosses a sharp 
molecular weight discontinuity.  Zoccali \etal\ (1999) have shown
that the luminosity of the RGB bump as
a function of metallicity as determined
from observation agrees well with that predicted by
the theory of stellar evolution.
Bono \etal\ (2001) further suggest that the agreement between the 
predicted
luminosity function and actual star counts along the RGB in the 
vicinity of
the bump in a suite of globular clusters is so
good that mixing cannot have occurred any earlier, otherwise the
evolutionary lifetimes, and hence the observed luminosity function, of 
such stars would have been affected by the mixing of He.

Zoccali \etal\ (1999) give the expected
location of the RGB bump in M5 to be 0.3 mag brighter
than the HB, i.e. at $V \sim14.8$ or $M_V \sim+0.5$.  Yet
we see strong
star-to-star variations in C and N abundances as well as a strong
anti-correlation between them within a
large group of cluster members
at $V \sim17.1~(M_V \sim+2.8)$, more than 2.3 mags fainter
at the base of the RGB.  We see hints of
such variation continuing on the upper main sequence at $M_V \sim+3.7$.

The range of luminosity over which these C and N
variations are seen is becoming more and more of a problem
for any scenario which invokes dredge up and mixing.
Unless we have missed some important aspect of
stellar evolution with impact on mixing and dredge up, we
must declare the mixing scenario a failure for the specific case
of M5 from our present work and M71 from our previous work
(and several other globular clusters as well from the work of others).
Even the theoreticians in the forefront of this field are beginning
to admit that deep mixing alone is not sufficient
(Denissenkov \& Weiss 2001, Ventura \etal\ 2001).
Unless and until some major new concept relevant to this issue appears,
we must now regard the fundamental origin of the star-to-star
variations we see in M5 as arising outside the stars whose
spectra we have studied here.

The strong anti-correlation between C and N, however, does
suggest that  CN-cycle material must be involved, and that
this material has somehow reached the surface
of these subgiant stars in M5.   Since we know it cannot come
from inside these stars, it must come from some external source.
As reviewed by Lattanzio, Charbonnel \& Forestini (1999),
CN and ON cycling is known
to occur in AGB stars, and AGB stars are also known to have sufficient
dredge up to bring such material to their surfaces.
We might speculate
that the site of the proton exposure could be a previous generation
of high mass stars, which then suffered extensive mass loss (either
in or outside of binary systems) and
polluted  the generation of lower-mass stars we currently observe,
while the higher mass stars are now defunct.

\subsection{ON Burning}

Let us adopt as a working hypothesis that the C and N abundance
variations we are seeing in the present
subgiants are the result of the incorporation of material exposed to
the CN-cycle (i.e., proton capture reactions) in a now evolved
population of more massive (2 - 5$M_{\odot}$) AGB stars, as was
originally suggested in D'Antona, Chieffi \& Gratton (1983).
Indeed, recent models of metal-poor AGB stars by Ventura \etal\ (2001)
suggest temperatures at the bases of the convective envelopes of such
stars are capable of these reactions.
The new generation of precision abundance analyses of globular
cluster stars over a wide range of luminosity such as that of
Ram\'{\i}rez \& Cohen (2002) for M71 demonstrate that the
abundances of the 
$\alpha$-capture and s-process elements are constant, and so we
further assume 
that the early cluster environment of M5 was
not significantly polluted by the ejecta of even more massive stars.

\begin{figure}
\epsscale{1.0}
\plotone{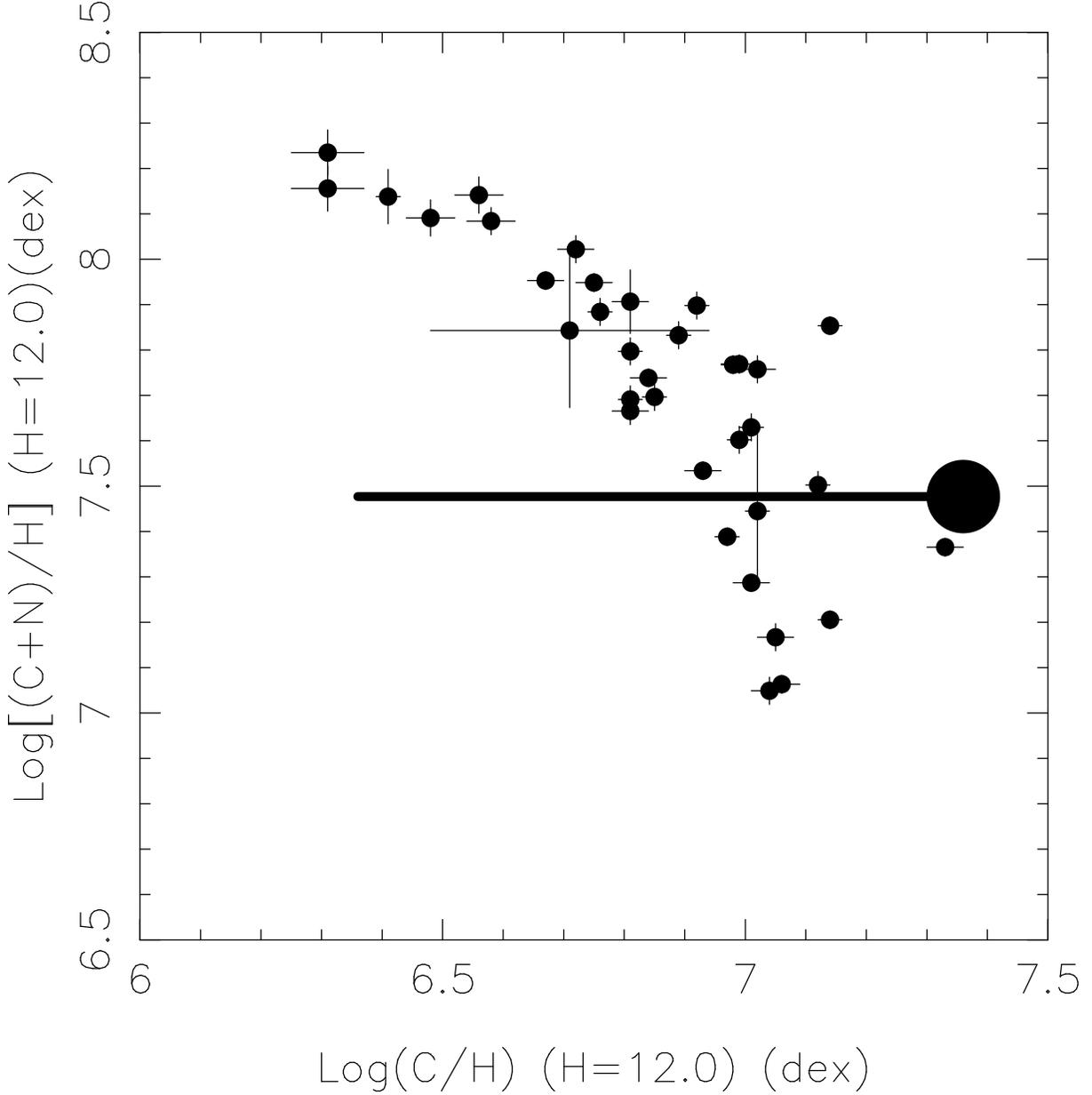}
\caption[jcohen_fig12.ps]{The sum of the derived C and N
abundances is plotted as a function of the C abundance.  The large
filled circle marks the location for both C and N depleted by a factor
of 16, adopting the abundance of M5 of [Fe/H] = $-1.2$ dex, with
C/N at the Solar ratio.  The horizontal line extending to the left
of that represents the locus of points for C gradually being converted
into N, with the left end of the line having C/C$_0$ = 0.1.
\label{fig_sum_cn}}
\end{figure}

We now investigate whether
this hypothesis is consistent with the C and N abundances we have
derived for the main subgiant sample of M5.  Figure~\ref{fig_sum_cn}
shows the sum of the C and N abundance as a function of the C
abundance of the sample of M5 subgiants.  The solid dot shows the
predicted location assuming the initial C and N  abundances
(C$_0$, N$_0$) are the Solar values reduced by the
metallicity of M5 ([Fe/H] = $-1.2$ dex).
Thus this is the initial location for no burning and for a Solar C/N
ratio.  If the present stars incorporated material in which just C was
burned into N, then the locus of the observed
points representing the M5 subgiant sample should consist of a
single horizontal line, with the initial point, the
presence of no CN-cycle exposed material, at the right end of the line
(the maximum C abundance) and the left end of the line corresponding
to a substantial fraction of the star's mass
(i.e. the atmosphere plus surface convection zone) 
including C-poor, N-rich AGB stellar ejecta.
Furthermore, if the initial C/N ratio of the cluster is not
Solar, then the locus should still be a horizontal line, but located
at a different vertical height in this figure.

The maximum possible N enhancement for a cluster SGB star with these
assumptions occurs if the star formed entirely from AGB ejecta 
in which
all C has been converted into N.
For initial values (C$_0$, N$_0$) (not expressed as logarithms),
this maximum N enhancement would be (C$_0$ + N$_0$)/N$_0$.
If the initial value was the Solar ratio, C$_0$/N$_0 \sim3.2$,
the resulting maximum N enhancement is a factor of $\sim$4.2,
while for an initial C$_0$/N$_0$ of 10, the maximum N enhancement
is a factor of 11.

Now we examine the behavior of the C and N abundances among
the M5 subgiant sample as inferred from our observations.
It is clear that the assumption that the only thing happening is
inclusion of material in which C was burned into N must be incorrect.
The sum of C+N seems to systematically increase 
by a factor of $\sim$5 between the
most C rich star and most C deficient star.
The discussion of the errors, both internal and systematic,
in \S\ref{section_abund} suggest maximum systematic errors
of $-0.2$ dex for log(C/H) and +0.2 for log(N/H).   This is completely
insufficient to explain such a large trend as errors.

We thus have a serious discrepancy.  The sum of C+N was
{\it{not}} constant as C was burned in the AGB sites.  Furthermore
the observed range in N abundances is very large.  The most 
obvious way to
reproduce this is to include O burning as well as C burning.
If we adopt Solar ratios as our initial values, then a substantial
amount of O burning is required.

Figure~\ref{fig_sum_cn} suggests that the
initial ratio of C/N is not quite
Solar, although not too far off.  Adopting the Solar value
as the initial C/N ratio, 
we calculate the minimum amount of O which
must be burned at the base of the AGB envelopes to reproduce the 
locus observed
in the figure (under the arguable assumption of the most extreme of
our stars having formed largely from such material - this will, however,
provide us with at least an estimate of the minimum burning required).
We need to produce a N enhancement of a factor of 10.
The Solar ratio is C/N/O = 3.2/1/7.6, so
if all the C and 50\% of the O were converted, we have an enhancement
of N of a factor of 8 available to the present stars.
Oxygen is typically found to be overabundant with respect to Fe in
old metal-poor systems (see Mel\'endez, Barbuy \& Spite 2001,
Gratton \etal\ 2001,
Ram\'{\i}rez \& Cohen 2002, and references therein); we assume
[O/Fe] $\sim +0.3$ dex, a typical value.
Then the initial C/N/O ratios will be 3.2/1/15.2.  Note that the 
same amount
of O has to be burned to produce the observed
distribution of C and N abundances,
but in this case it is a considerably
smaller fraction of the initial O.

Returning to the AGB models of Ventura \etal\ (2001), the
requirement for substantial O burning that emerges from
our analysis of the CH and CN bands in the M5 subgiants may not be an
unreasonable constraint - for metal-poor AGB stars they find 
temperatures
sufficient for CNO-processing at the bases of AGB stars in a wide mass
range. Following their Z=0.001 ([Fe/H] = --1.3) models we find surface O
abundances dropping by a factor of 2 to 20 in masses from 4.5 to 
2$M_{\odot}$.
We also note that under the assumption of little change in 
[C/Fe] (less than 0.3 dex as discussed above) taking place 
during the RGB ascent of the present
low mass M5 stars, one should also expect little change in [O/Fe] 
as well,
and that the O abundances of the present bright giants reflect their
``primordial'' values.  The observed [O/Fe] abundances of the bright M5
giants by Sneden \etal\ (1992) are not inconsistent with this idea. 
Their
``O-rich'' stars average around [O/Fe] = +0.3 while their most 
``O-poor'' stars are
depleted by a factor of 3.5. If this is the result of ON-cycle 
exposure, more
than enough N can be produced to explain the present results.

This simple test is of course leaves significant questions unanswered.
Problems include whether there were a sufficient number of AGB stars
present to return the required quantity of material and the 
efficiency of any
mechanism to incorporate it in the present stars. This is 
a non-trivial
issue if the most C/O-poor, N-rich SGB stars formed with a preponderance
of AGB ejecta - to reduce the C abundance by a factor of 6 by 
adding C-poor
ejecta would require some 83\% of the present star's mass to be 
made of this material.
Note, as has been pointed out by several authors, these
abundance inhomogeneities cannot simply be surface contaminations
as they would be diluted by the increasing depth of the convective 
envelope
during RGB ascent. Also, the range of C abundances among the M5 SGB
stars appears much larger than that of any cluster studied to date. 
While
this can perhaps be explained with regard to the more metal-rich 
clusters,
whose polluting AGB stars should have undergone less ON-cycling, other
even more metal-poor clusters than M5 appear to have smaller 
star-to-star
C variations among their less evolved stars. This can be seen in 
the [C/Fe]
versus luminosity diagrams of Carbon \etal\ (1982) and Bellman 
\etal\  (2001) for M92,
where the range in [C/Fe] among the least luminous stars is
relatively small (certainly
not the factor of 6 seen here in M5). But this may be perhaps 
explained 
by varying the efficiency for
incorporating AGB ejecta into subsequent generations of stars
among the proto-globular clusters.

Following Ventura \etal\ (2001), we
predict substantial O variations, anti-correlated with Na and Al, to be
present among the M5 subgiants and fainter stars. If the source of the proton
exposed materials is indeed moderate mass AGB stars, these 
inhomogeneities
should also correlate with Li abundances among the less evolved stars.
We expect the verification or lack thereof of these predictions
to be available shortly.

Another important point is that Figure~\ref{fig_sum_cn} shows no
evidence for
any bimodality in the distribution of the C and N
abundances  for the small section of the subgiant branch in M5
covered by our sample.  The distribution along the locus
of abundance appears to first order to be uniform.
There is no preponderance for stars populating the extremes
of high C or low C.  An artificial suggestion of bimodality, or
more correctly a tendency towards high CN strengths, could be
produced by the distribution of C and N abundances shown in
Figure~\ref{fig_c_vs_n} or by saturation effects in the
molecular bands themselves, which would only become apparent
in cooler stars with stronger molecular bands.

We also recall the anomalous stars in our sample, most of 
which are believed to
be members of M5.  Two of these in particular have enormously strong
CH bands; we can offer no explanation for these stars at present.

\subsection{Additional Implications for Stellar Evolution}

Figure~\ref{fig_sum_cn} was used above to demonstrate
that ON burning is required by considering the required
N enhancement factor.  This figure also displays a
correlation between the sum of C and N number densities with the 
C abundance, i.e. the C abundance is correlated with the C/N ratio
among the M5 subgiant stars. 
Here we explore the consequences 
of the existence of this correlation for
the origin of the C and N variations themselves.
Any external mechanism for producing these variations 
will involve an efficiency
factor for the incorporation of material.  We expect this factor
to depend on the mass of the
star itself,
how much additional mass is incorporated (${\Delta}M$),  
and the initial C and N abundances in the star itself and within ${\Delta}M$.
Since these properties of
${\Delta}M$ might be expected to fluctuate wildly, this
process therefore should show
a lot of stochastic random variability. 

It is easy to imagine
that various parcels of ${\Delta}M$ have a wide range in 
C/N ratios due to the
varying amount of nuclear processing that each might have experienced, 
and thus the strong stochastic star-to-star variations
in C and N as well as their overall anti-correlation that
we observe among the M5 subgiants
can be reproduced.  However, the correlation above requires
a correlation between the mass of the parcel
accreted, ${\Delta}M$, and the C/N ratio of
the material within this mass, and that seems
to be rather artificial given the random nature of the process.
We thus conclude that
is difficult to reproduce the correlation described above with such  
an external mechanism involving accretion of ``polluted''
material from AGB stars.  
Unless we have made gross errors 
in the C and N abundances in our M5 subgiant sample far beyond
what we believe might have occured, the existence of this correlation 
suggests that we should not totally rule out internal mechanisms as yet.

\section{Summary}

We have presented photometry and spectroscopy for a large sample of M5
subgiants and several main sequence turn off stars. 
An analysis of the SGB spectra reveals
significant and anti-correlated star-to-star variations in C and N 
abundances, as
would be expected from the presence of proton-capture exposed 
material in
our sample stars. Similar variations in CH also appear to be 
present in the
main sequence turn off spectra, but the signal in the 
current sample is too low for a detailed
analysis. The evolutionary states of these stars are such that the 
currently proposed
mechanisms for {\it in situ} modifications of C, N, O, etc. have 
yet to take place.  On this basis,
we infer that the source of proton exposure lies not within the 
present
stars, but more likely in a population of more 
massive (2 -- 5$M_{\odot}$) stars which
have ``polluted'' our sample. 

The C+N abundances derived for the M5
subgiants show a large systematic increase in C+N as C decreases.
To reproduce this requires  
the incorporation not only of CN, but of ON-processed material as 
well.  
Furthermore, the existence of this correlation is quite
difficult to reproduce with an external mechanism 
such as ``pollution'' with  material
processed in a more massive AGB star, which mechanism
is fundamentally stochastic in nature.  We therefore suggest
that although the internal mixing hypothesis has serious flaws,
new theoretical insights are needed and it should not be ruled 
out yet.

\acknowledgements

The entire Keck/HIRES and LRIS user communities owes a huge debt to
Jerry Nelson, Gerry Smith, Steve Vogt, Bev Oke, and many other
people who have worked to make the
Keck Telescope and HIRES and LRIS a reality and to operate and
maintain the Keck Observatory. We are grateful to the
W. M.  Keck Foundation for the vision to fund
the construction of the W. M. Keck Observatory.  The authors wish 
to extend
special thanks to those of Hawaiian ancestry on whose sacred mountain
we are privileged to be guests.  Without their generous hospitality,
none of the observations presented herein would
have been possible.

JGC  acknowledges support from the National Science Foundation 
(under grant
AST-9819614) and
MMB acknowledges support from the National Science Foundation 
(under grant
AST-0098489) and from the F. John Barlow endowed professorship. We are
also in debt to Roger Bell for the use of the SSG program and the 
Dean of the
UW Oshkosh College of Letters and Sciences for the workstation 
which made
the extensive modeling possible.

This work has made use of the USNOFS Image and Catalog Archive 
operated by
the United States Naval Observatory, Flagstaff Station
(http://www.nofs.navy.mil/data/fchpix/).

\clearpage

\clearpage

%
%
\begin{deluxetable}{lclll}
\tablenum{1}
\tablewidth{0pt}
\tablecaption{Probable Non-Members of M5}
\tablehead{
\colhead{ID} & \colhead{Sample} & \colhead{$v_r$} &
\colhead{Abs. Spectrum} & \colhead{$\Delta[V,I$]\tablenotemark{a}}\nl
\colhead{}
& \colhead{(P or S)} & \colhead{(\kms)} & \colhead{} & \colhead{(mag)}
}
\startdata
Non-members: \nl
C18262\_0726 &  P  & $-73.1$ & Metal rich & +0.03 \nl
C18465\_0208 & P & ...\tablenotemark{b} & Metal rich & +0.21 \nl
Probable Members: \nl
C18285\_0147 & P &  +60.7 & OK & +0.14 \nl
C18408\_0805 & P & ...\tablenotemark{b} & OK & $-$0.14 \nl
C18488\_0120 & P & ...\tablenotemark{b} & OK & $-$0.14 \nl
Uncertain: \nl
C18383\_0722\tablenotemark{c} & S &  ...\tablenotemark{b} & BHB ? & 
$-0.16$ \nl
\enddata
\tablenotetext{a}{This is [$V-I$ for the star] --  [$V-I$ for the cluster
locus] evaluated at the $V$ mag of the star.}
\tablenotetext{b}{No radial velocity is available for this star.}
\tablenotetext{c}{This faint star is possibly a blue straggler
in M5 or possibly a background early-type star.}
\label{table_nonmem}
\end{deluxetable}

%
%
\begin{deluxetable}{llll}
\tablenum{2}
\tablewidth{0pt}
\tablecaption{Photometry for Members of M5 in Our Sample}
\tablehead{
\colhead{ID\tablenotemark{a}} & \colhead{$B$} & \colhead{$V$} &
\colhead{$I$} \nl
\colhead{}
& \colhead{(mag)} & \colhead{(mag)} & \colhead{(mag}
}
\startdata
C18149\_0401 & 17.66 & 16.93 & 16.07 \nl
C18159\_0608 & 17.79 & 17.10 & 16.25 \nl
C18174\_0329 & 15.80 & 14.95 & 13.96 \nl
C18175\_0332 & 17.63 & 16.93 & 16.07 \nl
C18177\_0109 & 17.76 & 17.10 & 16.20 \nl
C18177\_0118 & 20.00 & 19.56 & 18.81 \nl
C18177\_0123 & 18.04 & 17.37 & 16.48 \nl
C18181\_0205 & 17.89 & 17.24 & 16.34 \nl
C18182\_0207 & 19.17 & 18.81 & 18.11 \nl
C18184\_0442 & 17.68 & 17.00 & 16.14 \nl
C18187\_0216 & 17.94 & 17.31 & 16.41 \nl
C18191\_0554 & 17.81 & 17.12 & 16.27 \nl
C18195\_0301 & 17.77 & 17.07 & 16.20 \nl
C18200\_0241 & 17.92 & 17.21 & 16.32 \nl
C18200\_0251 & 18.97 & 18.54 & 17.87 \nl
C18200\_0351 & 17.84 & 17.15 & 16.30 \nl
C18204\_0418 & 17.52 & 16.83 & 15.97 \nl
C18204\_0524 & 17.78 & 17.09 & 16.24 \nl
C18204\_0521 & 18.34 & 17.68 & 16.88 \nl
C18206\_0733 & 18.88 & 18.42 & 17.82 \nl
C18211\_0559 & 18.59 & 18.06 & 17.42 \nl
C18225\_0533 & 20.41 & 19.83 & 19.08 \nl
C18225\_0537 & 17.79 & 17.07 & 16.20 \nl
C18240\_0320 & 17.54 & 16.92 & 16.06 \nl
C18240\_0621 & 17.65 & 16.93 & 16.09 \nl
C18240\_0705 & 17.71 & 17.03 & 16.17 \nl
C18241\_0216 & 17.63 & 17.00 & 16.05 \nl
C18243\_0634 & 17.78 & 17.16 & 16.26 \nl
C18243\_0638 & 18.87 & 18.38 & 17.73 \nl
C18246\_0716  & 17.72 & 17.03 & 16.16 \nl
C18257\_0756 & 18.00 & 17.31 & 16.47 \nl
C18262\_0723 & 19.19 & 18.70 & 18.07 \nl
C18285\_0147 & 17.82 & 17.32 & 16.24 \nl
C18268\_0132 & 17.73 & 17.07 & 16.18 \nl
C18384\_0725 & 17.50 & 16.78 & 15.95 \nl
C18384\_0728 & 18.58 & 18.14 & 17.46 \nl
C18386\_0709 & 17.60 & 16.92 & 16.03 \nl
C18408\_0805 & 17.62 & 16.75 & 16.04 \nl
C18412\_0233 & 17.78 & 17.17 & 16.20 \nl
C18413\_0647 & 17.86 & 17.26 & 16.30 \nl
C18422\_0256 & 18.89 & 18.52 & 17.82 \nl
C18422\_0306 & 17.86 & 17.24 & 16.31 \nl
C18422\_0318 & 19.73 & (19.23)\tablenotemark{b} & 18.54 \nl
C18422\_0748 & 18.58 & 18.14 & 17.46 \nl
C18422\_0757 & 17.74 & 17.03 & 16.16 \nl
C18429\_0337 & 18.00 & 17.40 & 16.43 \nl
C18437\_0548 & 19.13 & 18.62 & 17.96 \nl
C18442\_0658 & 17.93 & 17.25 & 16.39 \nl
C18448\_0557 & 17.75 & 17.04 & 16.17 \nl
C18461\_0520 & 17.96 & 17.32 & 16.43 \nl
C18478\_0505 & 18.75 & 18.32 & 17.67 \nl
C18478\_0508 & 17.79 & 17.10 & 16.25 \nl
C18484\_0608 & 17.82 & 17.14 & 16.28 \nl
C18488\_0120 & 17.79 & 16.85 & 16.14 \nl
C18496\_0637 & 17.94 & 17.29 & 16.42 \nl
C18501\_0410 & 19.38 & 18.73 & 18.18 \nl
C18502\_0146 & 17.65 & 17.00 & 16.07 \nl
C18502\_0405 & 17.57 & 16.91 & 15.99 \nl
C18502\_0447 & 17.68 & 17.01 & 16.13 \nl
  ~ \nl
BHB Star \nl
C18386\_0713 & 15.56 & 15.60 & 15.59 \nl
\enddata
\tablenotetext{a}{The star names are derived from their
J2000 coordinates.  Star C12345\_5432 has coordinates
15 12 34.5~~+2 54 32 (J2000).}
\tablenotetext{b}{The $V$ mag is obtained from the $B,B-I$
assuming the object is on the main sequence of M5, as suggested
by the available $B,I$ photometry.}
\label{table_photmem}
\end{deluxetable}
\clearpage

%
%
\begin{deluxetable}{cccccccccccc}
\tablenum{3}
\tablewidth{0pt}
\tablecaption{Indices, Model Parameters, and Resulting Abundances}
\tablehead{
\colhead{Star} & \colhead{Comment} & \colhead{I(CH)} & 
\colhead{$\epsilon$I(CH)} & \colhead{S(3839)} & 
\colhead{$\epsilon$S(3839)\tablenotemark{a}} 
& \colhead{T$_{eff}$} & \colhead{log 
g} & \colhead{[C/Fe]} & \colhead{$\epsilon$[C/Fe]} & 
\colhead{[N/Fe]} & \colhead{$\epsilon$[N/Fe]}
}
\startdata
C18149\_0401 & SGB-CH-Strong & 0.192 & 0.004  & 0.236 & 0.012 & 
5192 & 3.23 & -0.35 & 0.02 & 0.66 & 0.03 \nl
C18159\_0608 & SGB-CH-Weak & 0.157 & 0.004  & 0.229 & 0.012 & 
5216 & 3.31 & -0.52 & 0.03 & 0.83 & 0.02 \nl
C18174\_0329 & SGB-CH-Strong & 0.221 & 0.004  & 0.346 & 0.012 & 
4849 & 2.29 & -0.42 & 0.01 & 0.58 & 0.03 \nl
C18175\_0332 & SGB-CH-Weak & 0.131 & 0.004  & 0.273 & 0.010 & 
5192 & 3.23 & -0.69 & 0.03 & 1.08 & 0.02 \nl
C18177\_0109 & SGB-CH-Weak & 0.152 & 0.004  & 0.191 & 0.012 & 
5216 & 3.31 & -0.55 & 0.03 & 0.75 & 0.03 \nl
C18177\_0118 & MS-CH-Weak & 0.022 & 0.012  & 0.049 & 
0.042 & - & - & - & - & - & - \nl
C18177\_0123 & Anomalous  & 0.193 & 0.004  & 0.015 & 0.016 & 5260 & 
3.44 & -0.30 & 0.03 & -1.87 & 0.02 \nl
C18181\_0205 & SGB-CH-Weak & 0.157 & 0.004  & 0.204 & 0.012 & 
5237 & 3.38 & -0.51 & 0.02 & 0.78 & 0.03 \nl
C18182\_0207 & MS-CH-Strong & 0.051 & 0.012  & 0.106 & 
0.044 & - & - & - & - & - & - \nl
C18184\_0442 & SGB-CH-Weak & 0.106 & 0.004  & 0.251 & 0.010 & 
5202 & 3.26 & -0.88 & 0.04 & 1.23 & 0.04 \nl
C18187\_0216 & SGB-CH-Strong & 0.169 & 0.004  & 0.160 & 0.012 & 
5249 & 3.41 & -0.43 & 0.03 & 0.56 & 0.02 \nl
C18191\_0554 & SGB-CH-Weak & 0.139 & 0.004  & 0.284 & 0.012 & 
5219 & 3.32 & -0.61 & 0.03 & 1.07 & 0.02 \nl
C18195\_0301 & SGB-CH-Weak & 0.136 & 0.004  & 0.232 & 0.012 & 
5212 & 3.30 & -0.65 & 0.23 & 0.96 & -0.17 \nl
C18200\_0241 & SGB-CH-Strong & 0.180 & 0.004  & 0.120 & 0.014 & 
5232 & 3.36 & -0.39 & 0.02 & 0.33 & 0.02 \nl
C18200\_0251 & MS-CH-Weak & 0.032 & 0.010  & 0.045 & 
0.032 & - & - & - & - & - & - \nl
C18200\_0351 & SGB-CH-Strong & 0.189 & 0.004  & 0.090 & 0.014 & 
5223 & 3.33 & -0.35 & 0.03 & 0.11 & 0.01 \nl
C18204\_0418 & SGB-CH-Weak & 0.102 & 0.004  & 0.255 & 0.010 & 
5178 & 3.18 & -0.95 & 0.02 & 1.28 & 0.06 \nl
C18204\_0521 & SGB-CH-Weak & 0.138 & 0.008  & 0.185 & 0.036 & 
5364 & 3.60 & -0.50 & 0.03 & 0.88 & 0.03 \nl
C18204\_0524 & SGB-CH-Weak & 0.152 & 0.004  & 0.244 & 0.012 & 
5215 & 3.31 & -0.55 & 0.02 & 0.90 & 0.03 \nl
C18206\_0533 & Anomalous - Strong CN/CH & 0.117 & 0.004  & 0.303 & 
0.012 & - & - & - & - & - & - \nl
C18211\_0559 & Anomalous - Strong CN/CH & 0.298 & 0.008  & 0.121 & 
0.044 & - & - & - & - & - & - \nl
C18225\_0537 & Anomalous  & 0.254 & 0.004  & 0.043 & 0.012 & 5212 & 
3.30 & -0.03 & 0.03 & -0.59 & 0.02 \nl
C18225\_0533 & MS-CH-Weak & 0.085 & 0.014  & -0.006 & 
0.062 & - & - & - & - & - & - \nl
C18240\_0320 & SGB-CH-Weak & 0.090 & 0.004  & 0.222 & 0.010 & 
5190 & 3.22 & -1.05 & 0.06 & 1.30 & 0.05 \nl
C18240\_0621 & SGB-CH-Strong & 0.212 & 0.004  & 0.410 & 0.010 & 
5192 & 3.23 & -0.22 & 0.02 & 0.91 & 0.02 \nl
C18240\_0705 & Anomalous  & 0.198 & 0.004  & 0.023 & 0.012 & 5206 & 
3.28 & -0.32 & 0.03 & -1.48 & 0.03 \nl
C18241\_0216 & SGB-CH-Weak & 0.145 & 0.004  & 0.272 & 0.010 & 
5202 & 3.26 & -0.60 & 0.02 & 1.00 & 0.03 \nl
C18243\_0634 & SGB-CH-Weak & 0.087 & 0.004  & 0.235 & 0.014 & 
5219 & 3.32 & -1.05 & 0.06 & 1.38 & 0.05 \nl
C18243\_0638 & MS-CH-Weak & 0.040 & 0.006  & 0.036 & 
0.020 & - & - & - & - & - & - \nl
C18246\_0716 & SGB-CH-Strong & 0.186 & 0.004  & 0.210 & 0.012 & 
5206 & 3.28 & -0.37 & 0.02 & 0.63 & 0.03 \nl
C18257\_0756 & SGB-CH-Weak & 0.149 & 
0.004 & - & -  & - & - & - & - & - & - \nl
C18262\_0723 & MS-CH-Weak & 0.020 & 
0.008 & - &  - & - & - & - & - & - & - \nl
C18268\_0132 & SGB-CH-Weak & 0.107 & 
0.004 & - & - & - & - & - & - & - & - \nl
C18285\_0147 & SGB-CH-Strong & 0.169 & 
0.004 & - & - &  - & - & - & - & - & - \nl
C18384\_0725 & SGB-CH-Strong & 0.175 & 0.004  & 0.357 & 0.014 & 
5170 & 3.16 & -0.44 & 0.02 & 1.00 & 0.03 \nl
C18384\_0728 & SGB-CH-Strong & 0.217 & 0.006  & 0.193 & 0.022 & 
5178 & 3.18 & -0.24 & 0.02 & 0.42 & 0.03 \nl
C18386\_0709 & SGB-CH-Strong & 0.186 & 0.004  & 0.303 & 0.016 & 
5190 & 3.22 & -0.37 & 0.03 & 0.84 & 0.02 \nl
C18408\_0805 & SGB-CH-Weak & 0.140 & 0.004  & 0.340 & 0.014 & 
5166 & 3.14 & -0.64 & 0.03 & 1.15 & 0.03 \nl
C18412\_0233 & SGB-CH-Strong & 0.165 & 0.004  & 0.285 & 0.012 & 
5226 & 3.34 & -0.47 & 0.02 & 0.93 & 0.03 \nl
C18413\_0647 & SGB-CH-Strong & 0.177 & 0.004  & 0.269 & 0.020 & 
5245 & 3.40 & -0.38 & 0.02 & 0.84 & 0.02 \nl
C18422\_0256 & MS-CH-Weak & 0.021 & 0.006  & 0.008 & 
0.020 & - & - & - & - & - & - \nl
C18422\_0306 & SGB-CH-Weak & 0.111 & 0.004  & 0.285 & 0.014 & 
5237 & 3.38 & -0.80 & 0.04 & 1.28 & 0.04 \nl
C18422\_0318 & MS-CH-Weak & 0.052 & 0.006  & 0.019 & 
0.018 & - & - & - & - & - & - \nl
C18422\_0748 & MS-CH-Weak & 0.023 & 0.010  & 0.041 & 
0.034 & - & - & - & - & - & - \nl
C18422\_0757 & SGB-CH-Strong & 0.201 & 0.004  & 0.052 & 0.022 & 
5206 & 3.28 & -0.31 & 0.03 & -0.31 & 0.03 \nl
C18429\_0337 & SGB-CH-Strong & 0.181 & 0.004  & 0.263 & 0.012 & 
5266 & 3.45 & -0.34 & 0.03 & 0.82 & 0.03 \nl
C18437\_0548 & MS-CH-Strong & 0.051 & 0.012  & 0.055 & 
0.040 & - & - & - & - & - & - \nl
C18442\_0658 & SGB-CH-Weak & 0.150 & 0.004  & 0.195 & 0.016 & 
5239 & 3.38 & -0.55 & 0.02 & 0.78 & 0.03 \nl
C18448\_0557 & Anomalous  & 0.218 & 0.004  & 0.043 & 0.016 & 5208 & 
3.28 & -0.22 & 0.02 & -0.50 & 0.02 \nl
C18461\_0520 & SGB-CH-Strong & 0.189 & 0.004  & 0.139 & 0.016 & 
5245 & 3.40 & -0.34 & 0.02 & 0.39 & 0.17 \nl
C18478\_0505 & MS-CH-Weak & 0.039 & 0.006  & -0.008 & 
0.018 & - & - & - & - & - & - \nl
C18478\_0508 & SGB-CH-Weak & 0.115 & 0.004  & 0.281 & 0.012 & 
5216 & 3.31 & -0.78 & 0.04 & 1.22 & 0.03 \nl
C18484\_0608 & SGB-CH-Weak & 0.149 & 0.004  & 0.283 & 0.014 & 
5222 & 3.33 &
  $-0.55$ & 0.03 & 1.02 & 0.07 \nl
C18488\_0120 & SGB-CH-Strong & 0.212 & 
0.004 & - & - & - & - & - & - &  - & - \nl
C18496\_0637 & SGB-CH-Weak & 0.108 & 
0.004 & - & - & - & - & - &  - & - & - \nl
C18501\_0410 & MS-CH-Strong & 0.047 & 
0.008 & - & - & - & - & - &  - & - & - \nl
C18502\_0146 & SGB-CH-Weak & 0.145 & 
0.004 & - & - & - & - & - &  - & - & - \nl
C18502\_0405 & SGB-CH-Strong & 0.164 & 
0.004 & - & - & - & - & - &  - & - & - \nl
C18502\_0447 & SGB-CH-Strong & 0.175 & 
0.004 & - & - & - & - & - &  - & - & -
\enddata
\tablenotetext{a}{The uncertainty in the zero point for S(3839) is
not included in the values listed above, and is $\pm$0.019.}
\label{table_obs_inds}
\end{deluxetable}
\clearpage

%
%
\begin{deluxetable}{cccccccccccccc}
\tablecolumns{14}
\tablenum{4}
\tablewidth{0pt}
\tablecaption{Model Parameters, Colors, and Resulting Indices}
\tablehead{
\colhead{} & \colhead{} & \colhead{} & \colhead{} & 
\multicolumn{2}{c}{[C/Fe]=$-$0.35, [N/Fe]=+0.10} & 
\multicolumn{2}{c}{[C/Fe]=$-$0.45, [N/Fe]=+1.0} &
\multicolumn{2}{c}{[C/Fe]=$-$0.55, [N/Fe]=+0.75} & 
\multicolumn{2}{c}{[C/Fe]=$-$0.65, [N/Fe]=+1.20} \nl
\colhead{T$_{eff}$}  & \colhead{log g} & \colhead{$M_V$} & 
\colhead{$V-I$} & \colhead{I(CH)} & \colhead{S(3839)} & 
\colhead{I(CH)} & \colhead{S(3839)} & \colhead{I(CH)} & 
\colhead{S(3839)}& \colhead{I(CH)} & \colhead{S(3839)} & 
\colhead{I(CH)} & \colhead{S(3839)}
}
\startdata
5103 & 2.94 & 1.887 & 0.851 & 0.225 & $-$0.001 & 0.189 & 0.278 & 
0.175 & 0.120 & 0.153 & 0.279 & 0.099 & 0.122 \nl
5159 & 3.12 & 2.288 & 0.836 & 0.217 & $-$0.010 & 0.182 & 0.243 & 
0.167 & 0.097 & 0.146 & 0.243 & 0.095 & 0.099 \nl
5216 & 3.31 & 2.689 & 0.821 & 0.208 & $-$0.018 & 0.174 & 0.209 & 
0.159 & 0.077 & 0.140 & 0.210 & 0.090 & 0.079 \nl
5291 & 3.50 & 3.092 & 0.802 & 0.196 & $-$0.029 & 0.164 & 0.168 & 
0.149 & 0.052 & 0.131 & 0.169 & 0.085 & 0.055 \nl
5379 & 3.61 & 3.293 & 0.778 & 0.181 & $-$0.042 & 0.151 & 0.120 & 
0.137 & 0.023 & 0.120 & 0.121 & 0.079 & 0.027 \nl
5626 & 3.79 & 3.500 & 0.714 & 0.131 & $-$0.069 & 0.109 & 0.006 & 
0.098 & $-$0.040 & 0.087 & 0.008 & 0.061 & $-$0.037 \nl
5997 & 3.99 & 3.696 & 0.625 & 0.070 & $-$0.094 & 0.061 & $-$0.078 & 
0.056 & $-$0.088 & 0.052 & $-$0.078 & 0.042 & $-$0.087 \nl
6148 & 4.11 & 3.889 & 0.592 & 0.056 & $-$0.102 & 0.050 & $-$0.094 & 
0.047 & $-$0.098 & 0.045 & $-$0.093 & 0.038 & $-$0.098 \nl
6209 & 4.21 & 4.089 & 0.580 & 0.052 & $-$0.104 & 0.047 & $-$0.097 & 
0.045 & $-$0.101 & 0.042 & $-$0.097 & 0.037 & $-$0.101 \nl
6214 & 4.28 & 4.287 & 0.581 & 0.053 & $-$0.103 & 0.047 & $-$0.096 & 
0.045 & $-$0.100 & 0.042 & $-$0.096 & 0.037 & $-$0.100 \nl
6186 & 4.35 & 4.488 & 0.589 & 0.055 & $-$0.100 & 0.049 & $-$0.092 & 
0.046 & $-$0.097 & 0.044 & $-$0.091 & 0.037 & $-$0.096 \nl
6133 & 4.41 & 4.648 & 0.589 & 0.061 & $-$0.095 & 0.053 & $-$0.084 & 
0.050 & $-$0.091 & 0.046 & $-$0.084 & 0.039 & $-$0.090 \nl
6062 & 4.46 & 4.892 & 0.618 & 0.070 & $-$0.089 & 0.060 & $-$0.074 & 
0.055 & $-$0.083 & 0.051 & $-$0.073 & 0.041 & $-$0.082 \nl
5980 & 4.51 & 5.094 & 0.636 & 0.082 & $-$0.083 & 0.069 & $-$0.060 & 
0.063 & $-$0.074 & 0.058 & $-$0.059 & 0.044 & $-$0.072 \nl
\enddata
\label{table_models}
\end{deluxetable}
\clearpage

%
%
\begin{deluxetable}{ccccccccccccc}\tablecolumns{13}
\tablenum{5}
\tablewidth{0pt}
\tablecaption{Changes in Derived C and N Abundances for Different 
Model Parameters}
\tablehead{
\colhead{}&\colhead{}&\colhead{}&\multicolumn{2}{c}{$\Delta$(m-
M)$_V$ = --0.10}&\multicolumn{2}{c}{[O/Fe] = +0.15} &
\multicolumn{2}{c}{\ciso = 4\tablenotemark{a}}
  & \multicolumn{2}{c}{[Fe/H] = --1.40}&\multicolumn{2}{c}{Turb = 
1.5 km/s} \nl
\colhead{Star}&\colhead{[C/Fe]}&\colhead{[N/Fe]}&\colhead{$\Delta$[C/Fe]}&\colhead{
$\Delta$[N/Fe]}&\colhead{$\Delta$[C/Fe]}&\colhead{$\Delta$[N/Fe]}&\colhead{
$\Delta$[C/Fe]}&\colhead{$\Delta$[N/Fe]} 
&\colhead{$\Delta$[C/Fe]}&\colhead{$\Delta$[N/Fe]}&\colhead{$\Delta$[C/Fe]
}&\colhead{$\Delta$[N/Fe]}
}
\startdata
C18200\_0351& --0.36 &  0.13  &      0.01     &       
0.01     &      0.03     &       0.01     &      0.00     &      
-0.02     &      0.08     &       0.11     &      0.05     &     
-0.07 \nl
C18429\_0337& --0.36 &  0.83  &      0.03     &       
0.01     &      0.05     &       0.00     &    --0.08     &       
0.06     &      0.09     &       0.14     &      0.05     &     
-0.04 \nl
C18243\_0634& --1.08 &  1.41  &      0.03     &       
0.00     &      0.04     &       0.01     &      0.01     &      
-0.02     &      0.14     &       0.07     &      0.11     &     
-0.11
\enddata
\tablenotetext{a}{Reduced from the adopted value of \ciso = 10.}
\label{table_changes}
\end{deluxetable}

\end{document}